\documentclass[10pt,twocolumn]{article}

\usepackage{amsmath,amstext,amsgen,amsbsy,amsopn,amsfonts,graphicx,overcite,
theorem,amssymb,psfrag}

\usepackage[nolists,noheads,nomarkers,tablesfirst]{endfloat}

\newcommand{\beq}{\begin{equation}}
\newcommand{\eeq}{\end{equation}}
\newcommand{\ee}[1] {\label{#1} \end{equation}}

\newcommand{\bea}{\begin{eqnarray}}

\newcommand{\eea}{\end{eqnarray}}

\newcommand{\bal}{\begin{align}}
\newcommand{\eal}{\end{align}}

\begin{document}

\title{A Perturbative Analysis of Modulated Amplitude Waves in Bose-Einstein 
Condensates}

\author{\\ Mason A. Porter \\ mason@math.gatech.edu \\ School of Mathematics 
and Center for Nonlinear Science \\ Georgia Institute of Technology, Atlanta,
 Georgia  30332 
\\ \\
Predrag Cvitanovi\'c \\ School of Physics and Center 
for Nonlinear Science \\ Georgia Institute of Technology, Atlanta, Georgia  
30332 \\ }

\date{\today}
\maketitle

\begin{abstract}

        We apply Lindstedt's method and multiple scale perturbation theory to
 analyze spatio-temporal structures in nonlinear Schr\"odinger equations and 
thereby study the dynamics of quasi-one-dimensional Bose-Einstein condensates
 with mean-field interactions.  We determine the dependence of the amplitude 
of modulated amplitude waves on their wave number.  We also explore the band 
structure of Bose-Einstein condensates in detail using Hamiltonian 
perturbation theory and supporting numerical simulations. 

\end{abstract}

\vspace{.1 in}


\vspace{.2 in}

{\bf Bose-Einstein condensates (BECs) were observed experimentally in 1995 
using dilute vapors of sodium and rubidium.  The macroscopic behavior of BECs 
at zero temperature is modeled by the nonlinear Schr\"odinger equation in the 
presence of an external potential.  This model has proven to be an excellent 
one for most experiments on BECs.  When the external potential is spatially 
periodic (e.g., due to an optical lattice, which may be created using 
counter-propagating laser beams), the spectrum of the BEC exhibits a band 
stucture (spatial resonance structure).  This paper utilizes Hamiltonian 
perturbation theory and supporting numerical simulations to study this 
structure in detail.}

\section{Introduction}

        At low temperatures, particles in a dilute gas can reside in the same
 quantum (ground) state, forming a Bose-Einstein 
condensate.\cite{pethick,stringari,ketter,edwards}  This was first observed 
experimentally in 1995 with vapors of rubidium and sodium.\cite{becrub,becna}
  In these experiments, atoms were confined in magnetic traps, evaporatively 
cooled to tempuratures on the order of fractions of microkelvins, left to 
expand by switching off the confining trap, and subsequently imaged with 
optical methods.\cite{stringari}  A sharp peak in the velocity distribution 
was observed below a critical temperature, incidating that 
Bose-Einstein condensation had occurred.

        BECs are inhomogeneous, so condensation can be observed in both 
momentum and coordinate space.  The number of condensed atoms $N$ ranges from
 several thousand to several million.  Confining traps are usually 
approximated well by harmonic potentials.  There are two characteristic 
length scales: the harmonic oscillator length 
$a_{ho} = \sqrt{\hbar/(m\omega_{ho})}$ [which is on the order of a few 
microns], where $\omega_{ho}=(\omega_x \omega_y \omega_z)^{1/3}$ is the 
geometric mean of the trapping frequencies, and the mean healing length 
$\chi=1/\sqrt{8\pi |a| \bar{n}}$,where $\bar{n}$ is the mean density and $a$, 
the (two-body) $s$-wave scattering length, is determined by the atomic 
species of the condensate.\cite{pethick,stringari,kohler,baiz}  Interactions 
between atoms are repulsive when $a > 0$ and attractive when $a < 0$.  For a 
dilute ideal gas, $a \approx 0$. The length scales in BECs should be 
contrasted with those in systems like superfluid helium, in which the effects 
of inhomogeneity occur on a microscopic scale fixed by the interatomic 
distance.\cite{stringari}

        If considering only two-body, mean-field interactions, a dilute 
Bose-Einstein gas can be modeled using a cubic nonlinear Schr\"odinger 
equation (NLS) with an external potential, which is also known as the 
Gross-Pitaevskii (GP) equation.  BECs are modeled in the 
quasi-one-dimensional (quasi-1D) regime when the transverse dimensions of the
 condensate are on the order of its healing length and its longitudinal 
dimension is much larger than its transverse 
ones.\cite{bronski,bronskirep,bronskiatt,stringari}  In the quasi-1D regime, 
one employs the 1D limit of a 3D mean-field theory rather than a true 1D 
mean-field theory, which would be appropriate were the tranverse dimension on 
the order of the atomic interaction length or the atomic 
size.\cite{bronski,bronskiatt,bronskirep,salasnich,towers}  

        When examining only two-body interactions, the condensate 
wavefunction (``order parameter'') $\psi(x,t)$ satisfies a cubic NLS,
\begin{equation}
        i\hbar\psi_t = -[\hbar^2/(2m)]\psi_{xx} + g|\psi|^2\psi + V(x)\psi 
\,, \label{nls3}
\end{equation}
 where $|\psi|^2$ is the number density, $V(x)$ is an external potential, 
$g = [4\pi\hbar^2 a/m][1 + \mathcal{O}(\zeta^2)]$, and 
$\zeta = \sqrt{|\psi|^2|a|^3}$ is the dilute gas 
parameter.\cite{stringari,kohler,baiz}  Because the scattering length $a$ can
 be adjusted using a magnetic field in the vicinity of a Feshbach 
resonance\cite{fesh}, the contribution of the nonlinearity in (\ref{nls3}) is
 tunable.

        Potentials $V(x)$ of interest in the context of BECs include harmonic
 traps, periodic potentials (``standing light waves''), and periodically 
perturbed harmonic traps.  The existence of quasi-1D cylindrical 
(``cigar-shaped'') BECs motivate the study of periodic potentials without a 
confining trap along the dimension of the periodic lattice.\cite{band}  
Experimentalists use a weak harmonic trap on top of the periodic lattice to 
prevent the particles from spilling out.  To achieve condensation, the 
periodic lattice is typically turned on after the trap.  If one wishes to 
include the trap in theoretical analyses, $V(x)$ is modeled by
\begin{equation}
        V(x) = V_0\sin(\kappa (x-x_0)) + V_1x^2 \,, \label{harmwiggle}
\end{equation}
where $\kappa$ is the lattice wave number, $V_0$ is the height of the 
periodic lattice, and $x_0$ is the offset of the periodic potential.  (Note 
that these three quantities can all be tuned experimentally.)  The periodic 
term dominates for small $x$, but the harmonic trap otherwise becomes quickly
 dominant.  When $V_1 \ll V_0$, the potential is dominated by its periodic 
contribution for many (20 or more) periods.\cite{kutz,promislow,lattice}  
(For example, when $V_0/V_1 = 500$, $\kappa = 10$, and $x_0 = 0$, the 
harmonic component of $V(x)$ essentially does not contribute for 10 periods.)
  In this work, we usually let $V_1 = 0$ and focus on periodic potentials.  
Spatially periodic potentials have been employed in experimental studies of 
BECs\cite{hagley,anderson} and have also been studied theoretically.\cite{bronski,bronskiatt,bronskirep,space1,space2,promislow,kutz,malopt,alf,smer}  

When the optical lattice has deep wells (large $|V_0|$), an analytical 
description of BECs in terms of Wannier wave functions can be obtained in the
 tight-binding approximation.\cite{smer2}  The Bose-Hubbard Hamiltonian, 
which is a better description than (\ref{nls3}) in the tight-binding 
approximation, is derived by expanding the field operator in a Wannier basis 
of localized wave functions at each lattice site.  This Hamiltonian has has 
three contributions: a kinetic energy term yielding contributions from 
tunnelling between adjacent wells, an energy offset in each lattice site 
(due, for example, to external confinement), and a potential energy term 
characterized by atom-atom interactions (that indicates how much energy it 
takes to put a second atom into a lattice site that already has one atom 
present).  One can use the Bose-Hubbard Hamiltonian to examine transitions 
between superfluidity and Mott insulation.\cite{mott}

In the present paper, we examine in detail the band structure of BECs in 
shallow periodic lattices using Hamiltonian perturbation theory and 
supporting numerical simulations.\cite{mapbecprl}  Our methodology, which 
yields analytical expressions describing the features of BEC resonance bands,
 exploits the elliptic function solutions of the NLS in the absence of a 
potential.  Note, however, that this paper does {\it not} explore the chaotic
 dynamics of BECs.

\section{Coherent Structures}

        We consider uniformly propagating coherent structures with the ansatz 
$\psi(x-vt,t) = R(x - vt)\exp\left(i\left[\theta(x-vt) - \mu t\right]\right)$,
where $R \equiv |\psi|$ is the magnitude (amplitude) of the wave function, 
$v$ is the velocity of the coherent structure, $\theta(x)$ determines its 
phase, $\vec{v}_0 \propto \nabla \theta$ is the particle velocity, and $\mu$ 
is the chemical potential (which can be termed an angular frequency from a 
dynamical systems perspective).  Considering a coordinate system that travels
 with speed $v$ (by defining $x' = x - vt$ and relabeling $x'$ as $x$) yields
\begin{equation}
        \psi(x,t) = R(x)\exp\left(i\left[\theta(x) - \mu t\right]\right) 
\,.  \label{maw2}
\end{equation}
[From a physical perspective, we consider the case $v = 0$, as 
$V(x') = V(x-vt)$.]  When the (temporally periodic) coherent structure 
(\ref{maw2}) is also spatially periodic, it is called a 
{\it modulated amplitude wave} (MAW).\cite{lutz1,lutz2}  The orbital stability
 of MAWS for the cubic NLS with elliptic potentials has been studied by 
Bronski and co-authors.\cite{bronski,bronskiatt,bronskirep}  To obtain 
stability information about the sinusoidal potentials we consider, one takes 
the limit as the elliptic modulus $k$ approaches zero.\cite{lawden,rand} 

When $V(x)$ is periodic, the resulting MAWs generalize the Bloch modes that
 occur in the theory of linear systems with periodic potentials, as one is 
considering a nonlinear Floquet-Bloch theory rather than a linear 
one.\cite{675,ashcroft,band,space1,space2}  In this paper, we employ phase 
space methods and perturbation theory to examine MAWs and their concomitant 
band structure.

The novelty of our work lies in its illumination of BEC band structure 
through the use of perturbation theory and supporting numerical simulations 
to examine $2m'\!:\!1$ spatial subharmonic resonances in BECs in periodic 
lattices.  Such resonances correspond to spatially periodic solutions $\psi$ 
of period $2m'$ and generalize the `period doubled' states (in $|\psi|^2$) 
studied by Machholm, {\it et al.}\cite{pethick2} which pertain to the 
experiments of Cataliotti, {\it et al.}.\cite{cata}  

Previous theoretical work in this area has focused on different aspects of 
BEC band structure, such as loop structure\cite{diak,machholm,wu4} and 
hysteresis.\cite{mueller}  In contrast to the coherent structures we 
consider, these authors studied band structure using a Bloch wave ansatz.  In
 our notation, they assumed {\it a priori} that $R(x) = R(x + 2\pi/\kappa)$ 
has the same periodicity of the underlying spatial lattice $V(x)$, whereas we
 have made no such assumption and instead use Hamiltonian perturbation theory
 to study the dynamical behavior of $R(x)$.  Additionally, the analytical 
components of these works are confined to two-to-three Fourier mode 
truncations of the Bloch wave dynamics.\cite{diak,machholm,wu4}

Inserting (\ref{maw2}) into the NLS (\ref{nls3}) and equating 
real and imaginary parts yields
\begin{align}
        \hbar\mu R(x) &= -\frac{\hbar^2}{2m}R''(x) \notag \\
        &\quad + \left[\frac{\hbar^2}{2m}\left[\theta'(x)\right]^2 + gR^2(x) + V(x) \right]R(x) \,, \\
        0 &= \frac{\hbar^2}{2m}\left[2\theta'(x)R'(x) + \theta''(x)R(x)\right] 
             \,, \notag
\end{align}
which gives the following two-dimensional system of nonlinear 
ordinary differential equations:
\begin{align}
        R' &= S \,, \notag \\
        S' &= \frac{c^2}{R^3} - \frac{2m\mu R}{\hbar} 
+ \frac{2mg}{\hbar^2}R^3 + \frac{2m}{\hbar^2}V(x)R 
\,. \label{dynam35}
\end{align}
The parameter $c$ is defined via the relation
\begin{equation}
        \theta'(x) = \frac{c}{R^2} \,,  \label{angmom}
\end{equation}
and therefore plays the role of ``angular momentum,'' as 
discussed by Bronski and coauthors.\cite{bronski}  [Equation (\ref{angmom}) is
 a statement of conservation of angular momentum.]  Constant phase solutions, 
which constitute an important special case, satisfy $c = 0$.

\begin{figure}
                \centerline{
                (a)
                \includegraphics[width=0.4\textwidth, height=0.5\textwidth]
{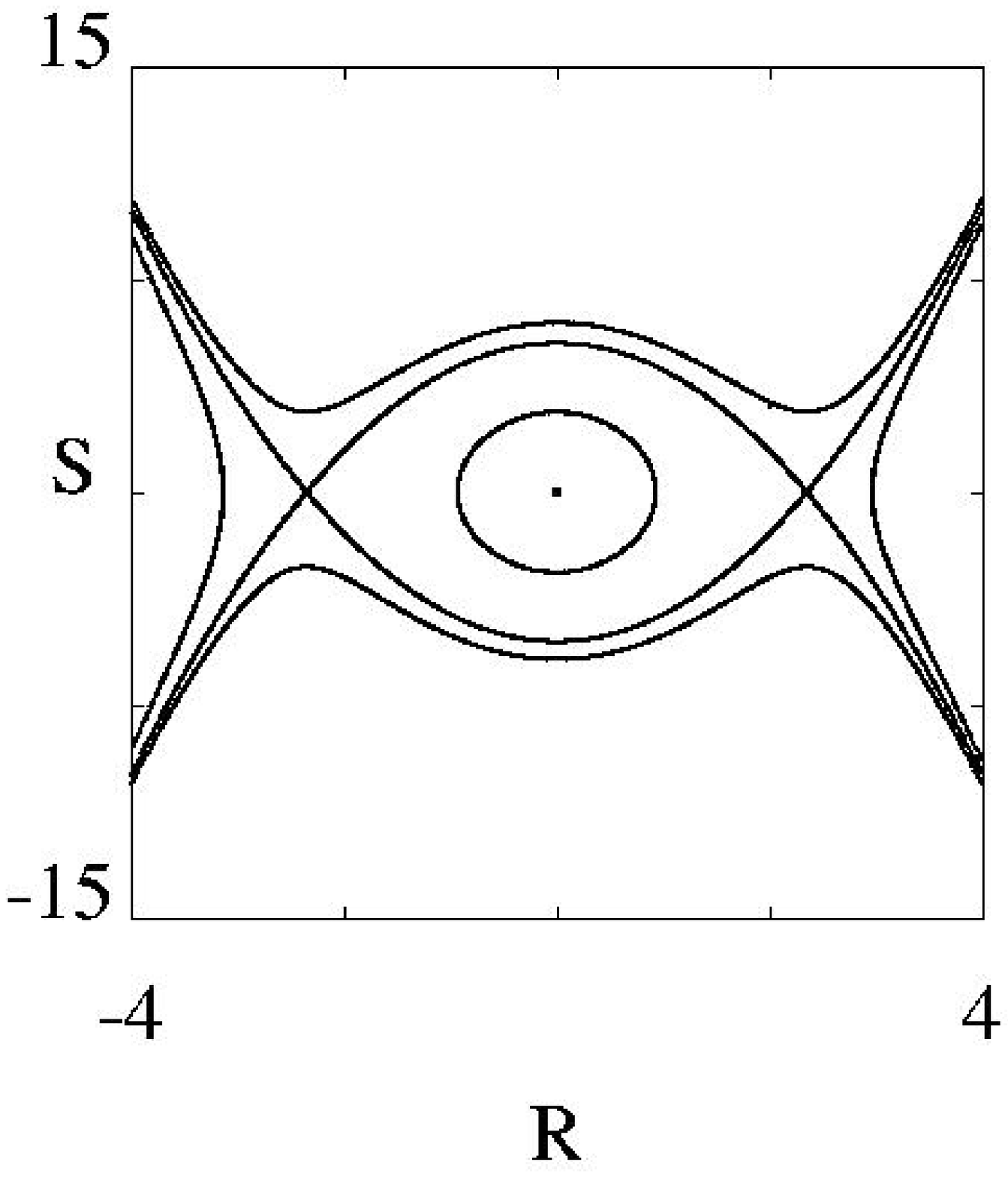}
                \hspace{1 cm}
                (b)
                \includegraphics[width=0.4\textwidth, height=0.5\textwidth]
{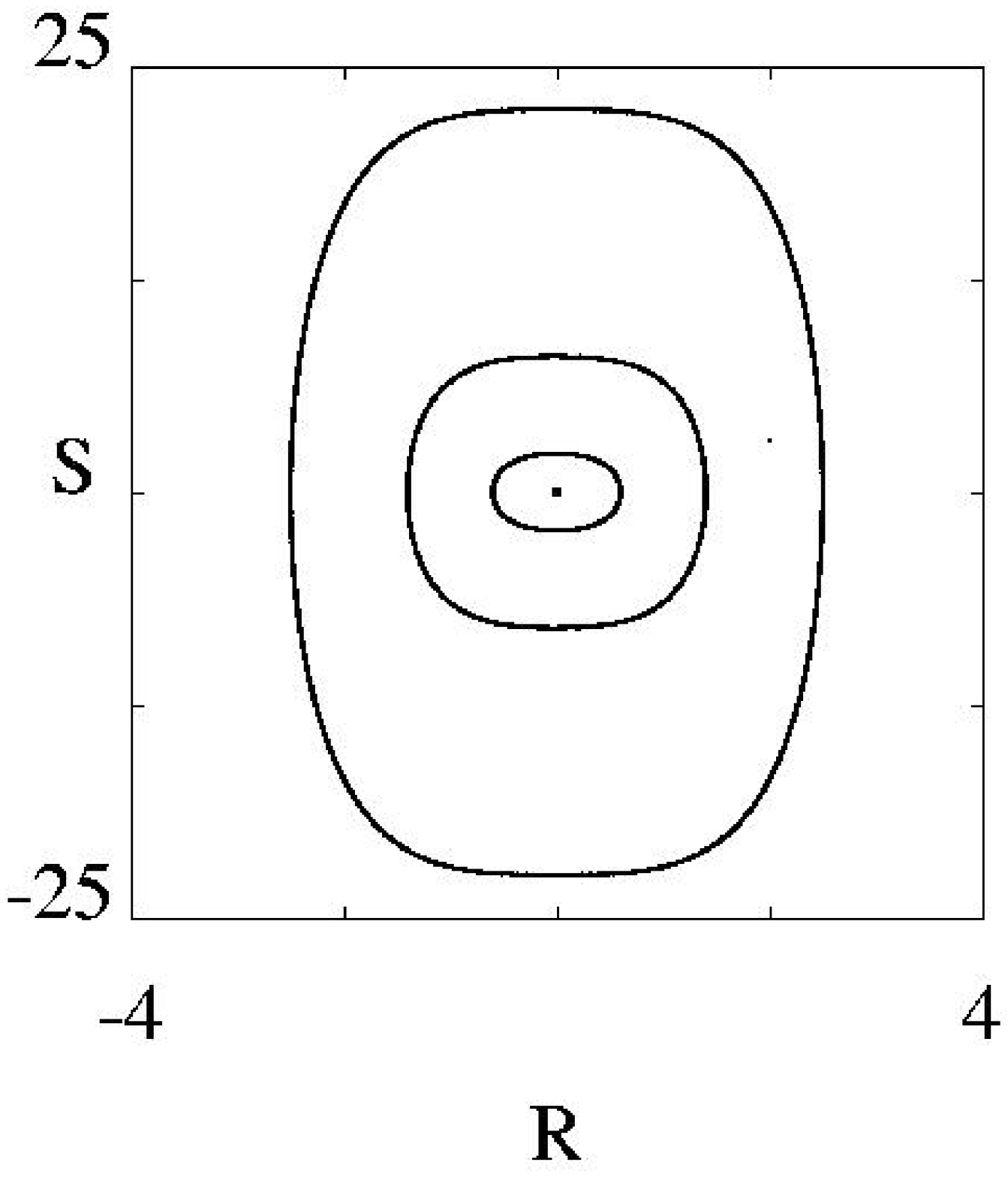}}
                \centerline{
                (c)
                \includegraphics[width=0.4\textwidth, height=0.5\textwidth]
{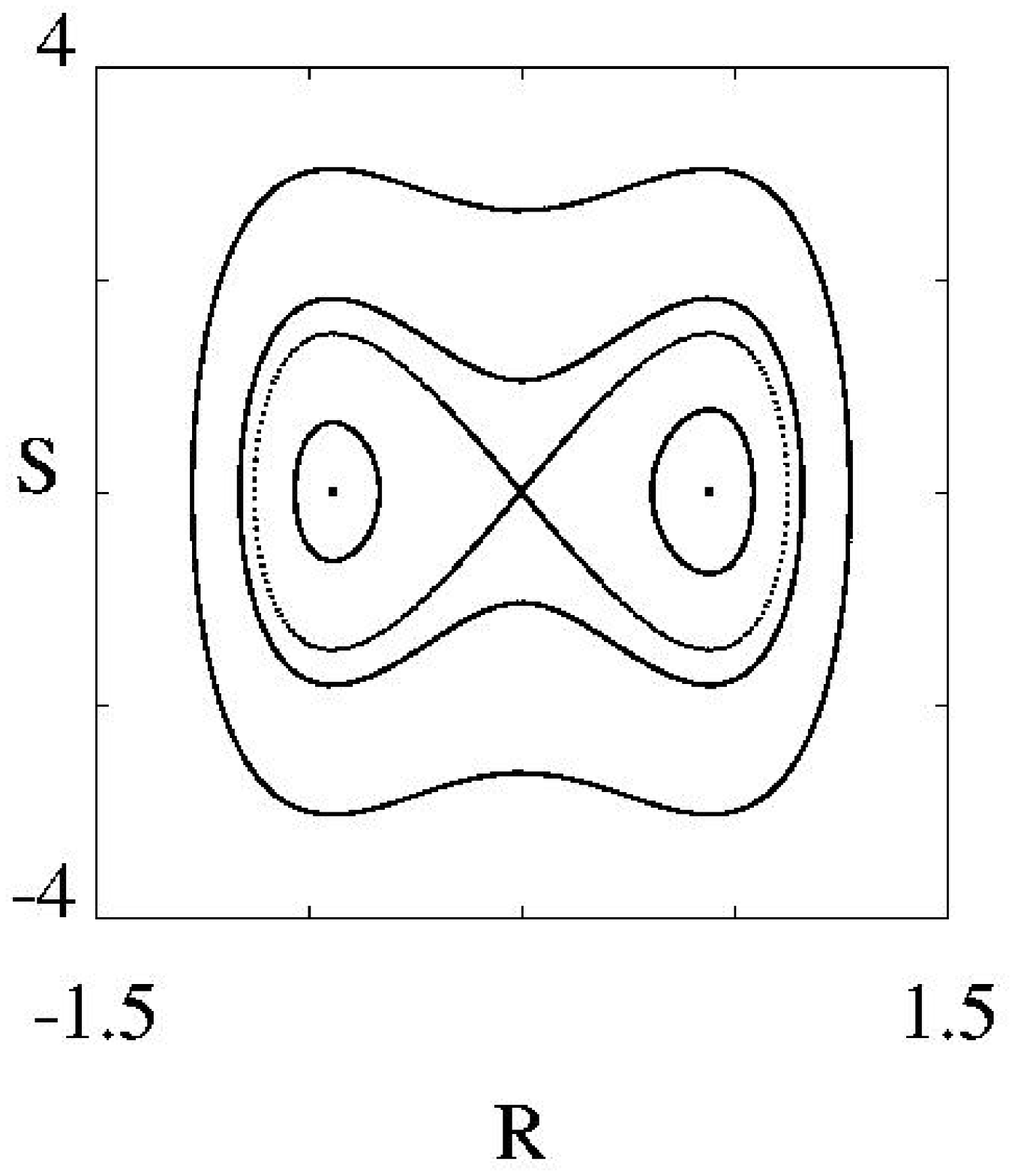}}


                \caption{Phase portraits of coherent structures in BECs with 
no external potential.  The signs of $\mu$ and $g$ determine the dynamics 
of (\ref{dynam35}).  (a) Repulsive BEC with $\mu > 0$.  The two-body 
scattering length is $a = 0.072$ nm, the value\cite{fried} for atomic 
hydrogen ($^1$H).  Orbits inside the separatrix (which consists of two 
heteroclinic orbits) have bounded amplitude $R(x)$.  The period of such 
orbits increases as one approaches the separatrix, whose period is infinite. 
 (b) Attractive BEC with $\mu > 0$.  The two-body 
scattering length is $a = -0.9$ nm, the value\cite{rub85,kutz} for $^{85}Rb$.
 (c) Attractive BEC (again $^{85}Rb$) with $\mu < 0$.  Here there are two 
separatrices, each of which encloses periodic orbits satisfying $R \neq 0$.} 
\label{repulse1}

\end{figure}

\section{BECs without an External Potential}

        When $V(x) \equiv 0$, the dynamical system (\ref{dynam35}) is 
autonomous and hence integrable, as it is two-dimensional.  Its equilibria 
$(R_*,S_*)$ satisfy $S_* = 0$ and either $R_* = 0$, $c = 0$ or  
\begin{equation}
        gR^6 - \hbar\mu R^4 + \frac{\hbar^2}{2m}c^2 = 0 \,, 
\label{equil35}
\end{equation}
which can be solved exactly because it is cubic in $R^2$.  When $c = 0$, one
 obtains $R_* = \pm \sqrt{\hbar\mu/g}$.  One thus obtains equilibria at 
$(R_*,0) \neq (0,0)$ for $g > 0$ if $\mu >0$ and $g < 0$ if $\mu < 0$.  

The eigenvalues of the equilibrium $(R_*,0)$ satisfy
\begin{align}
        \lambda^2 = -\frac{3c^2}{R_*^4} - \frac{2m\mu}{\hbar} 
+ \frac{6mg}{\hbar^2}R_*^2 \,.  
\label{vals35}
\end{align}
When $c = 0$ and $R_* = 0$, one obtains 
$\lambda = \pm \sqrt{-2m\mu/\hbar}$.  Additionally, one obtains a 
center at $(0,0)$ when $\mu > 0$ and a saddle when $\mu < 0$.  One also
 obtains saddles at the $R_* \neq 0$ equilibria for $g > 0$ when 
$\mu > 0$ and centers at those same locations for $g < 0$ when 
$\mu < 0$.  These latter equilibria are surrounded by periodic orbits 
that satisfy $R \neq 0$.  The possible qualitative dynamics (for $c = 0$) 
are illustrated in Figure \ref{repulse1} and summarized in Table \ref{tab2}.

\begin{table}[t] 
\centerline{
\begin{tabular}{|c|c|c|r|} \hline
Equilibrium at $(0,0)$ & Equilibria at $(R_*,0) \neq (0,0)$ & $g$ & $\mu$
 \\ \hline
Center & None   & $-$   & $+$ \\
Center & None   & $0$   & $+$ \\
Center & Saddles   & $+$   & $+$ \\
Saddle & None   & $+$   & $-$ \\
Saddle & None   & $0$   & $-$ \\
Saddle & Centers   & $-$   & $-$ \\ \hline
\end{tabular}}\caption{Type of equilibria of (\ref{dynam35}) when $V(x) = 0$,
 and $c = 0$.}\label{tab2}
\end{table}

\begin{figure}
                \centerline{
                (a)
                \includegraphics[width=0.32\textwidth, height=0.40\textwidth]
{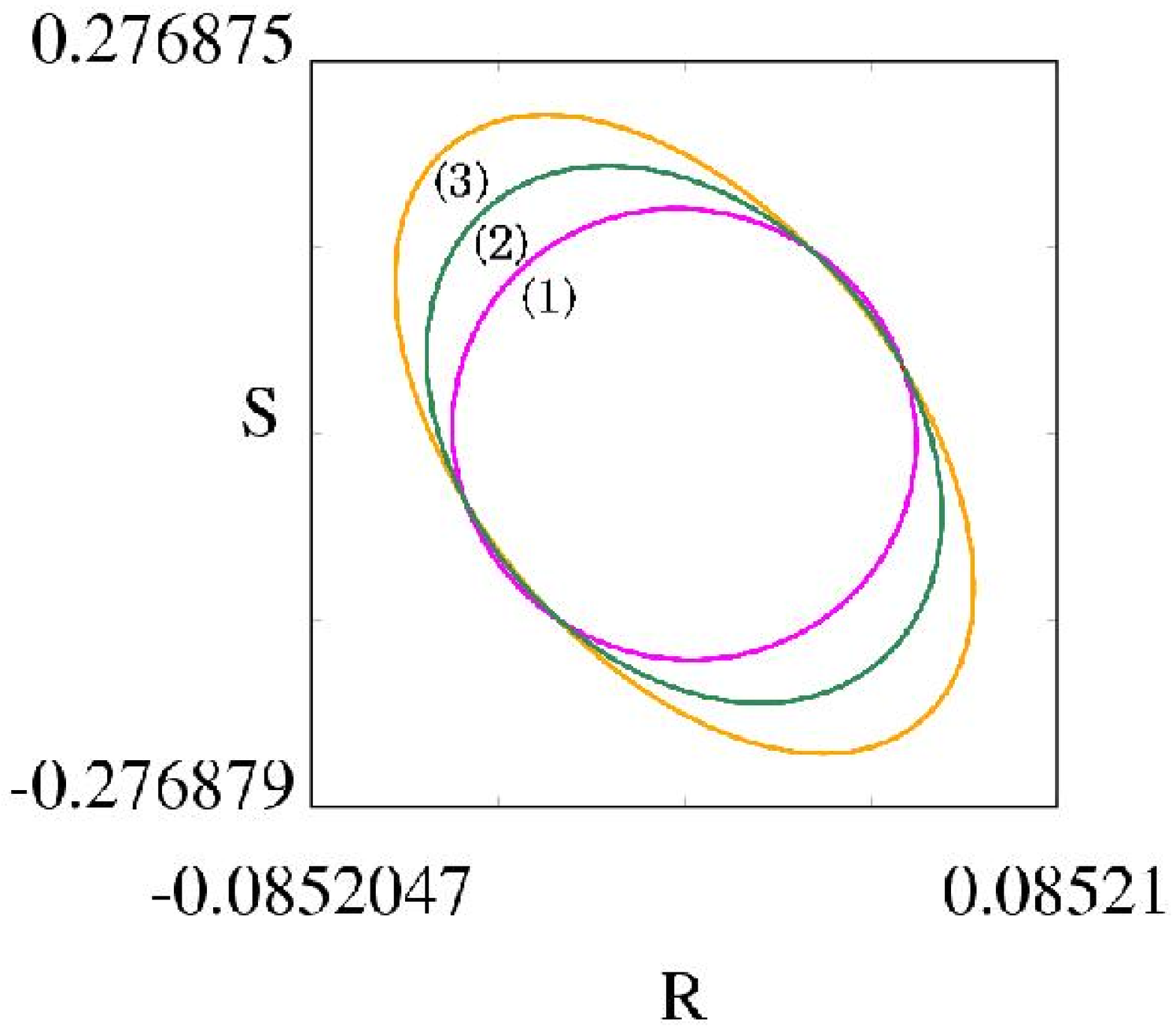}
                \hspace{1 cm}
                (b)
                \includegraphics[width=0.32\textwidth, height=0.40\textwidth]{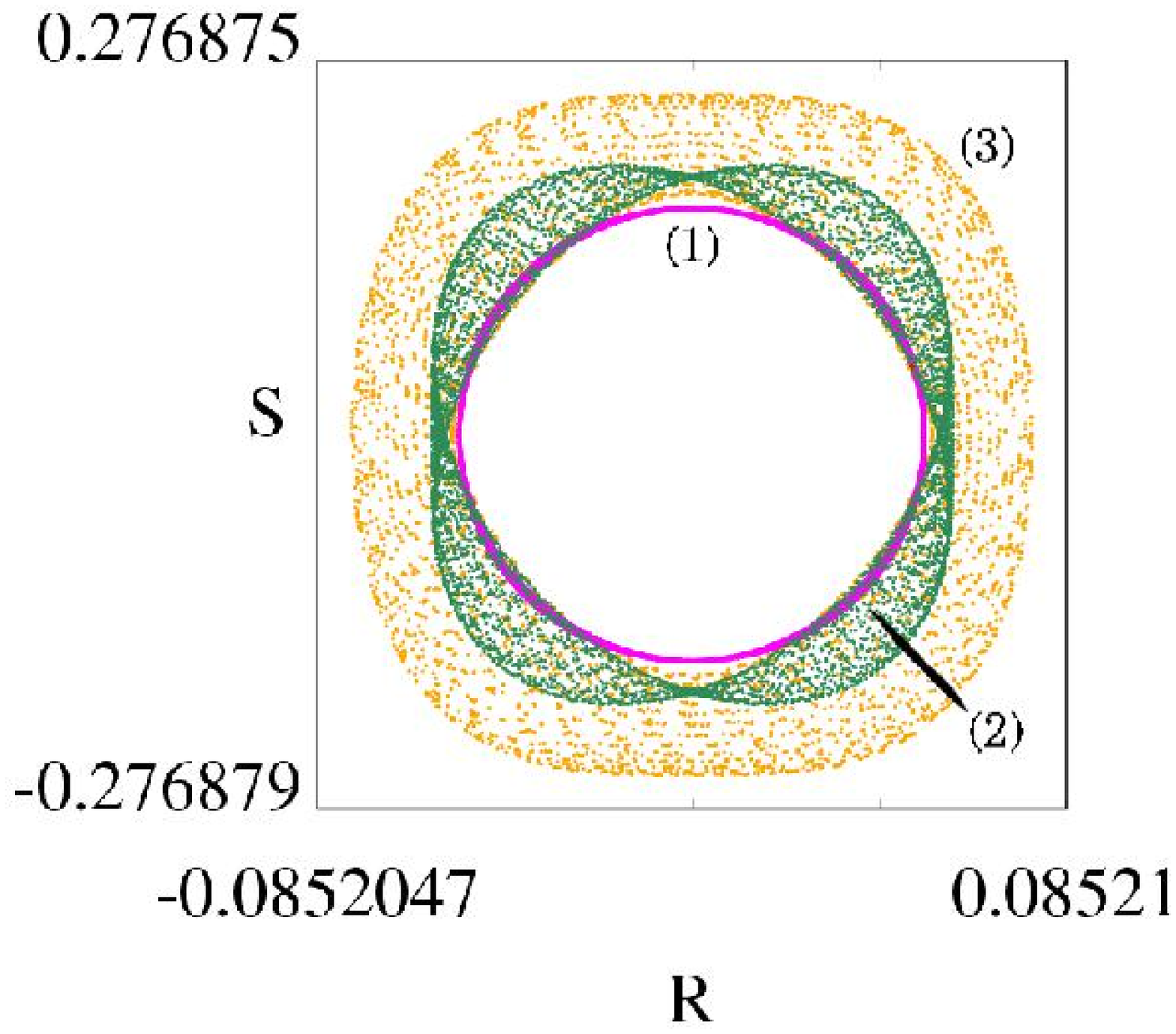}}
                \centerline{
                (c)
                \includegraphics[width=0.32\textwidth, height=0.40\textwidth]{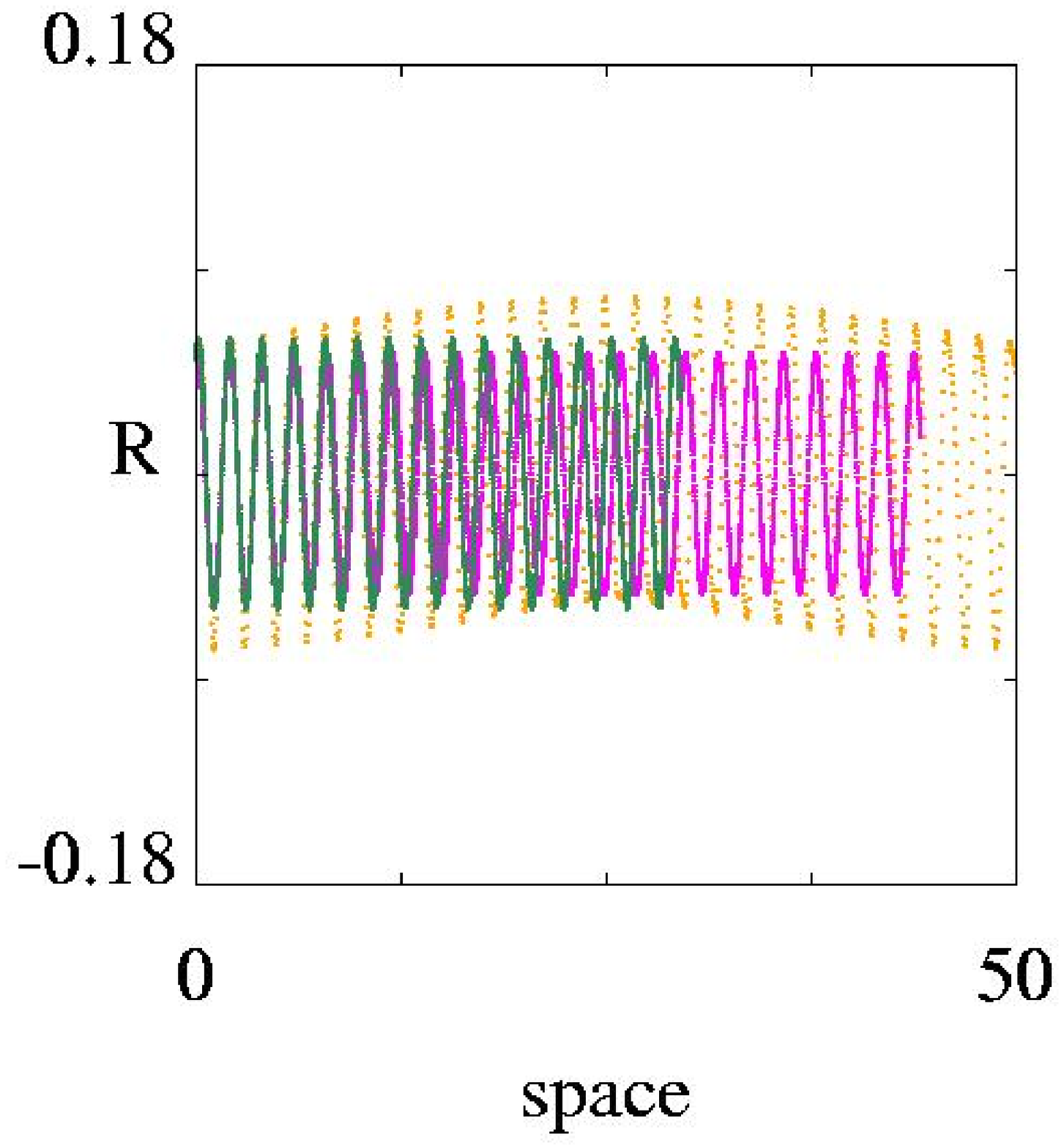}
                \hspace{1 cm}
                (d)
                \includegraphics[width=0.32\textwidth, height=0.40\textwidth]{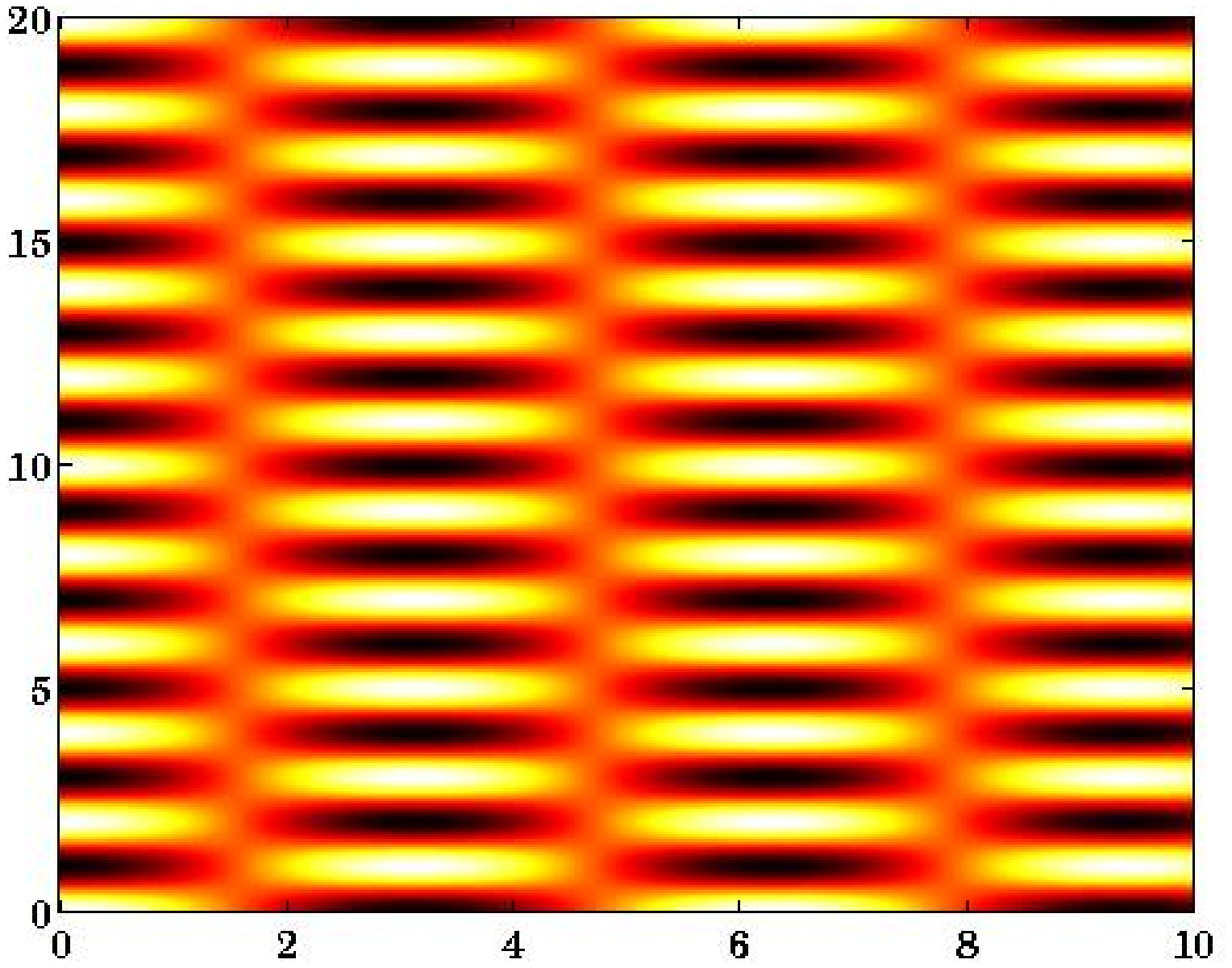}}

                \caption{As the wave number $\kappa$ of the perturbation is 
increased, periodic behavior persists for larger $|V_0|$.  The initial
 condition in this plot is $(R(0),S(0)) = (0.05,0.05)$, and the parameter 
values $a = 0.072$, $\mu = 10$, $m = 0.5$, and $x_0 = 0$ are used for each
 trajectory.  (a) Poincar\'e section determined by $\sin(\kappa x) = 0$. 
 Trajectory (1) corresponds to $(\kappa,V_0) = (100,10)$, trajectory (2) to 
$(\kappa,V_0) = (100,100)$, and trajectory (3) to $(\kappa,V_0) = (10,10)$.  
These quasiperiodic solutions indicate the existence of nearby periodic orbits.
  (b) Phase space plots of the trajectories in (a).  Trajectory (1) is the 
closest to being periodic and trajectory (3) is the furthest away.  (c) 
Amplitude $R$ as a function of space $x$ for trajectories (1)--(3).  The band 
structure of BECs can be studied not only in real space but also in phase 
space by plotting Poincar\'e sections and trajectories, as indicated in (a) 
and (b).  Examining the proximity of a trajectory to periodicity is most 
easily accomplished in phase space.  (d) Coherent structure corresponding to
 quasiperiodic trajectory (1).  This plot depicts $Re(\psi)$.  The horizontal
 axis represents time, and the vertical one represents space.  The darkest 
portions are the most negative, and the lightest are the most positive.} 
\label{kaphy}

\end{figure}

\begin{figure}
        \begin{centering}
                \includegraphics[width = 2.5 in, height = 3 in]
{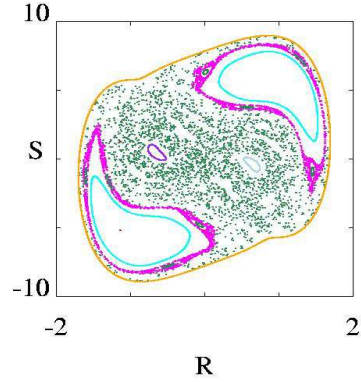}


                \caption{Poincar\'e section for the parameter values 
$\mu = -10$, $m = 0.5$, $x_0 = 0$, $V_0 = 5$, $\kappa = 10$, and 
$a = -0.9$ nm, corresponding to the experimentally determined scattering 
length\cite{rub85,kutz} for $^{85}$Rb.  The depicted trajectories include 
examples which are quasiperiodic, locally chaotic (near the resonances), and 
globally chaotic (the stochastic sea).} \label{rb}
        \end{centering}
\end{figure}

\begin{figure}
                \centerline{
                (a)
                \includegraphics[width=0.4\textwidth, height=0.5\textwidth]
{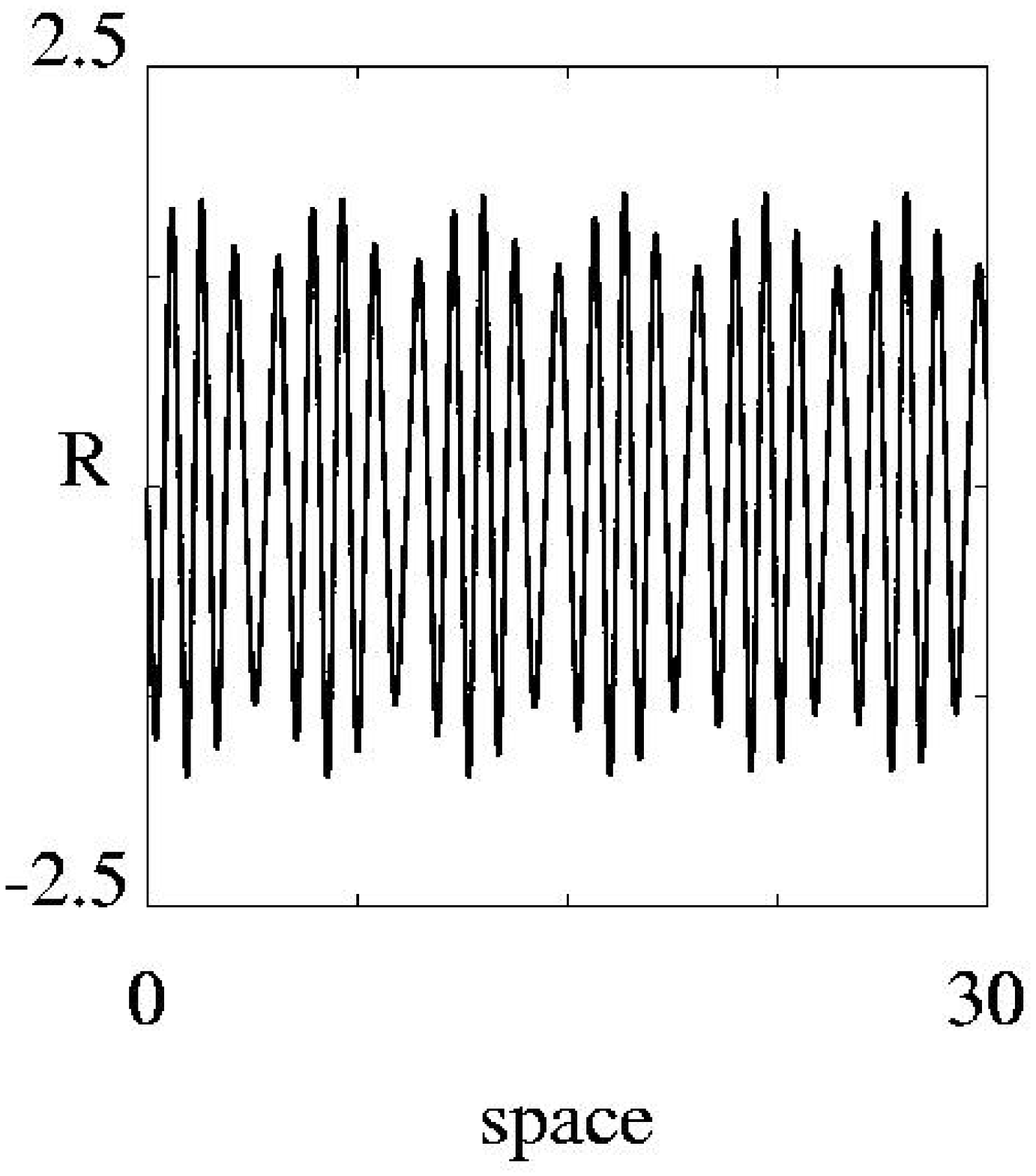}
                (b)
                \includegraphics[width=0.4\textwidth, height=0.5\textwidth]
{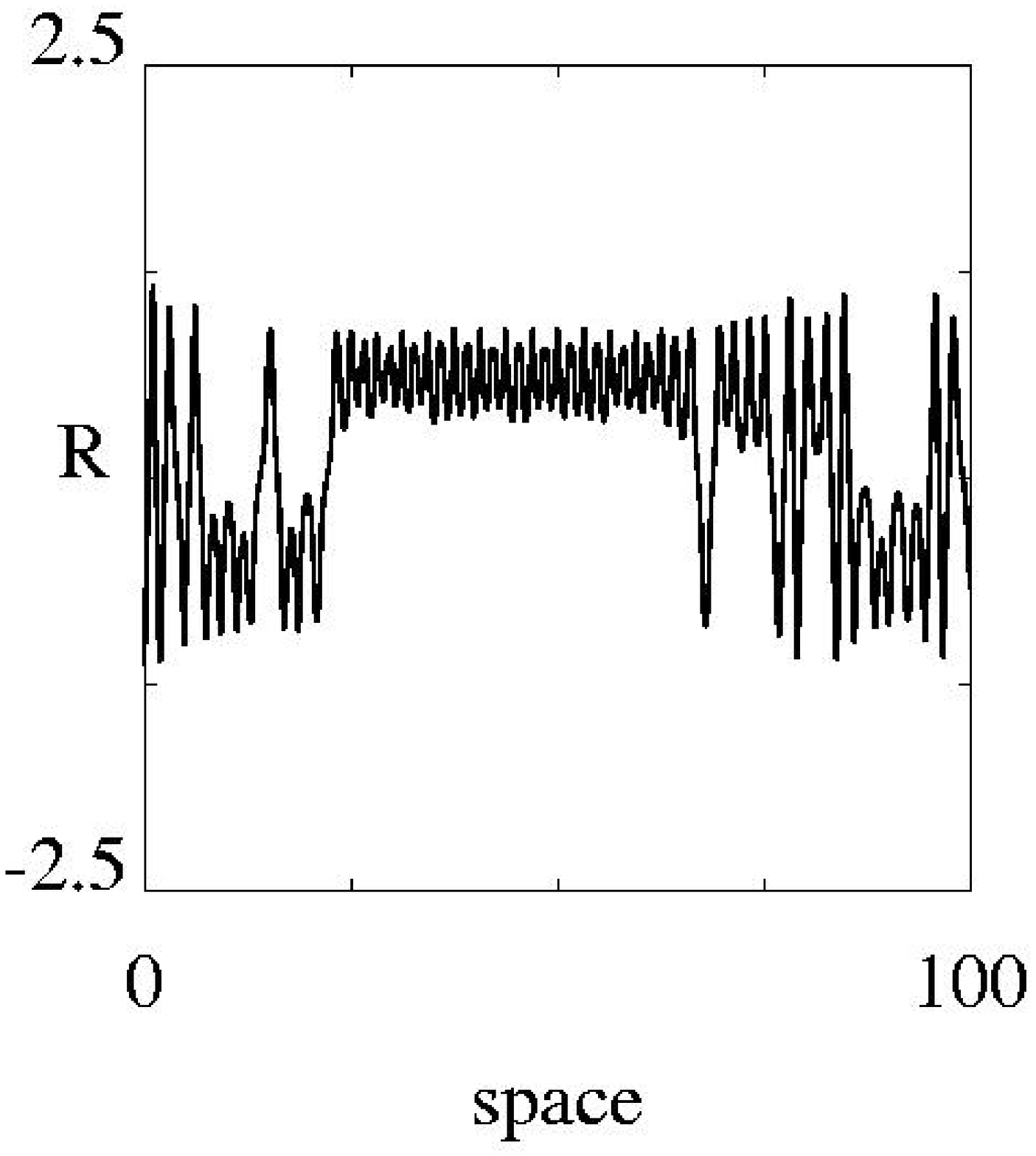}}


                \caption{(a) Spatial profile of the coherent structure 
corresponding to the locally chaotic trajectory in Figure \ref{rb}.  The 
initial conditions are $(R(0),S(0)) \approx (-0.01818215,-5.23268358)$.  (b) 
Spatial profile of the coherent structure corresponding to the globally 
chaotic trajectory in Figure \ref{rb}.  The initial conditions are 
$(R(0),S(0)) \approx (-1.13283530,1.28334013)$.} \label{time}

\end{figure}

To study the dependence of the wave number of periodic orbits (centered at 
the origin) of (\ref{dynam35}) on the amplitude $R$ when $V(x) \equiv 0$, we 
employ Lindstedt's method\cite{675} and consider null angular-momentum 
wave functions for the case $\mu > 0$.  We also assume 
$g = \varepsilon \bar{g}$, where $\varepsilon \ll 1$ and 
$\bar{g} = \mathcal{O}(1)$.  The period of $R(x)$ is given by
\begin{equation}
        T = \frac{2\pi}{\alpha} 
= 2\pi\left[1 + \frac{3gA^2}{8\mu\hbar}\right] 
+ \mathcal{O}(\varepsilon^2) \,,  \label{ampy}
\end{equation}
where $R(\xi) = R_0(\xi) + \mathcal{O}(\varepsilon)$, $\xi := \alpha x$, 
$\alpha = 1 + \varepsilon \alpha_1 + \mathcal{O}(\varepsilon^2)$ is the wave 
number, $A := R_0(0)$, and
\begin{equation}
        R_0(\xi) = A\cos\left(\sqrt{\frac{2m\mu}{\hbar}}\xi\right) \,.
\end{equation}

Note that all periodic orbits are centered at the origin when $\mu > 0$.  
When $g < 0$, the spatial period becomes smaller with increasing $A$.  When 
$g > 0$, the period becomes larger with increasing $A$.  In the latter case, 
the wave number-amplitude relation holds only for solutions inside the 
separatrix, as the trajectories are unbounded outside the separatrix and 
hence not periodic.  

Before deriving the wave number-amplitude relations when $V(x) \neq 0$, we 
comment briefly on the preceeding results.  The spatial period for small 
$c \neq 0$ is similar to (\ref{ampy}), but it cannot be estimated as easily 
because equation (\ref{dynam35}) now includes a term of order 
$\mathcal{O}(R^{-3})$ with coefficient $c$.  Although (\ref{ampy}) can be 
computed exactly in terms of elliptic functions, here we are 
interested in elucidating the qualitative dynamics of the MAWs of interest as
 well as establishing the methodology to be employed in the presence of 
potentials $V(x)$.  We will utilize elliptic function solutions in Section 
\ref{subharm} in our detailed study of band structure.\cite{mapbecprl}  The 
physical relevance of elliptic functions to BECs has been discussed by Carr 
and collaborators.\cite{rein1,rein2}

\section{BECs in a Periodic Lattice}
  
To study the wave number-amplitude relations of periodic orbits in the 
presence of external potentials, we expand the spatial variable $x$ in 
multiple scales.  We define ``stretched space'' $\xi := \alpha x$ as in the 
integrable situation and ``slow space'' $\eta := \varepsilon x$.  We consider 
potentials of the form $V(x) = \varepsilon \bar{V}(\xi,\eta)$, where 
$\bar{V}(\xi,\eta) = \bar{V}_0\sin\left[\kappa(\xi-\xi_0)\right] 
+ \bar{V}_1(\eta)$  and $\bar{V}_1$, which is of order $\mathcal{O}(1)$, is 
arbitrary but slowly varying.  Cases of particular interest include 
$\bar{V}_1 = 0$ (periodic potential) and 
$\bar{V}_1 = \tilde{V}_1(\eta - \eta_0)^2$ (superposition of periodic and 
harmonic potentials).  

When $\kappa \neq \pm 2\beta := \pm 2\sqrt{2m\mu/\hbar}$, the 
equations of motion for the slow dynamics of (\ref{dynam35}) with $c = 0$
 are
\begin{align}
        \frac{dA}{d\eta} \equiv A' &= -\beta\alpha_1B 
- \frac{3\beta\bar{g}}{8\mu\hbar}BC^2 
- \frac{\beta}{2\mu\hbar}B\bar{V}_1(\eta) \,, \notag \\
        \frac{dB}{d\eta} \equiv B' &= \beta\alpha_1A 
+ \frac{3\beta\bar{g}}{8\mu\hbar}AC^2 
+ \frac{\beta}{2\mu\hbar}A\bar{V}_1(\eta) \,, \label{slow35}
\end{align}
where $g \equiv \varepsilon \bar{g}$.  The leading-order expression for the 
amplitude is
\begin{equation}
        R_0(\xi,\eta) = A(\eta)\cos(\beta\xi) + B(\eta)\sin(\beta\xi) \,, 
\label{sol0}
\end{equation}
where $C^2 := A(\eta)^2 + B(\eta)^2$ is a constant.  The dynamical system 
(\ref{slow35}) is autonomous when $V_1(x) \equiv 0$.  Equilibria 
$(A_*,B_*) \neq (0,0)$ of (\ref{slow35}) correspond to periodic orbits of 
(\ref{dynam35}) with $c = 0$.  The equilibrium value of the squared amplitude 
is denoted $C_*^2 = A_*^2 + B_*^2$.

Converting to polar coordinates with $A(\eta) = C\cos(\phi(\eta))$ and 
$B(\eta) = C\sin(\phi(\eta))$ and integrating the resulting equation yields
\begin{equation}
        \phi(\eta) = \phi(0) 
+ \left[\alpha_1\beta + \frac{3\beta\bar{g}}{8\mu\hbar}C^2\right]\eta 
+ \frac{\beta}{2\mu\hbar}\int\bar{V}_1(\eta)d\eta \,.
\end{equation}

The wave number of the periodic motion is given by
\begin{equation}
        \alpha(C) = 1 - \frac{3g}{8\mu\hbar}C^2 
- \frac{1}{2\mu\hbar}V_1(x) 
+ \mathcal{O}(\varepsilon^2) \,.  \label{alf35}
\end{equation}

When $\kappa = \pm 2\beta$, we show that the slow flow equations have an 
extra term due to resonance.  Without loss of generality, we let 
$\kappa = +2\beta$, as changing the sign of $V_0$ produces the 
$\kappa = -2\beta$ case.  When $m = 0.5$, $\mu = 10$, and $\hbar = 1$, for
 example, one obtains this resonant situation for $\kappa = \pm 10$.  
Additionally, we show that $C$ is no longer constant in this resonant 
situation.  

In polar coordinates, the slow flow equations are
\begin{align}
        \phi' &= \alpha_1\beta + \frac{3\beta\bar{g}}{8\mu\hbar}C^2 
+ \frac{\bar{V}_0\beta}{4\mu\hbar}\sin[2(\phi - \beta\xi_0)] 
+ \frac{\beta}{2\mu\hbar}\bar{V}_1(\eta) \,, \notag \\
        C' &= -\frac{\bar{V}_0\beta}{4\mu\hbar}C\cos[2(\phi - 2\beta\xi_0)] 
\,. \label{slowb35}
\end{align}
Integrating the equation for $C'$ yields
\begin{equation}
        C = C_0 \exp\left[-\frac{\bar{V}_0\beta}{4\mu\hbar}\int\cos[2(\phi(\eta)-\beta\xi_0)]d\eta\right], 
\end{equation}
which one may then insert into the equation for the angular dynamics.

To determine equilibria, one puts $C' = \phi' = 0$.  From $C' = 0$, one 
determines that equilibria $(C_*,\phi_*)$ satisfy
\begin{equation}
        \phi_* = \frac{(2j+1)\pi}{4} + \beta\xi_0 \,, \hspace{.1 in} 
j \in \{0,1,2,3\}\,, 
\label{angeq}
\end{equation}
which is independent of the scattering coefficient.  Inserting (\ref{angeq}) 
into $C' = 0$ yields the wave number-amplitude relation
\begin{equation}
        \alpha_R(C) = \alpha(C) \mp \frac{V_0}{4\mu\hbar} 
+ \mathcal{O}(\varepsilon^2) \,, \label{perper35}
\end{equation}
for periodic orbits of (\ref{dynam35}).  In (\ref{perper35}), the minus sign 
is obtained when $j \in \{0,2\}$, and the plus sign is obtained when 
$j \in \{1,3\}$.  Equation (\ref{perper35}) is valid for
 $2\!:\!1$ spatial resonances.  We examine $2m'\!:\!1$ resonances for integer
 $m'$ in Section \ref{subharm} using Hamiltonian perturbation theory and the 
elliptic function solutions of (\ref{dynam35}) when $V_0 = 0$.

To examine the spatial stability (i.e., stability with respect to spatial 
evolution) of these periodic orbits in the presence of resonant periodic 
potentials, we compute the spatial stability of equilibria of (\ref{slowb35}) 
when $V_1(x) \equiv 0$.  The eigenvalues of the periodic orbits are 
\begin{equation}
        \lambda = \pm \frac{\beta}{2\mu}\sqrt{\mp\frac{3\bar{V}_0\bar{g}C}{2\hbar^2}} \,.
\end{equation}
We show numerical simulations for (\ref{dynam35}) in the presence of a 
periodic potential in Figure \ref{kaphy}.  In this situation, 
(\ref{dynam35}) is a nonlinear Mathieu equation.\cite{675,rand,zounes}  
Figure \ref{kaphy}d shows the coherent structure for the trajectory with 
$\kappa = 100$ and $V_0 = 10$.  Figure \ref{rb} depicts a Poincar\'e section 
describing the dynamics of $^{85}Rb$, for which $a = -0.9$ nm.  Figure 
\ref{time} depicts spatial profiles of the coherent structures corresponding 
to the locally chaotic and globally chaotic trajectories in Figure \ref{rb}.

\section{Subharmonic Resonances} \label{subharm}

        In this section, we analyze spatial subharmonic resonances and the 
band structure of repulsive BECs with a positive chemical potential.  We 
perturb off the elliptic function solutions of the underlying integrable 
system in order to study $2m'\!:\!1$ spatial resonances with a leading-order 
perturbation method.  Perturbing off simple harmonic functions, by contrast, 
requires a perturbative method of order $m'$ to study $2m'\!:\!1$ resonances.
  At the center of the KAM islands, we observe `period-multiplied' states.
  When $m' = 1$, one obtains period-doubled states in $\psi$.  As verified 
numerically in Section \ref{numsect}, our qualitative results are 
excellent.  Given that our method is a leading-order one, our quantitative 
results are also remarkably good.

        Recent work by Machholm and coauthors\cite{pethick2} on 
period-doubled states (in $|\psi|^2$) follows up experimental studies by 
Cataliotii and coauthors,\cite{cata} who observed superfluid current 
disruption in chains of weakly coupled BECs, which is related to the 
dependence of the dynamical instability of Bloch states on the magnitude of 
particle interactions.  Period-doubled states, which may be interpreted as 
soliton trains, arise from dynamical instabilities of the energy 
bands associated with Bloch states.\cite{pethick2}  In the present work, we 
offer a dynamical 
systems perspective on period-doubled states and their generalizations.  Our 
theoretical and computational analysis reveals period-multiplied solutions of 
the GP (\ref{nls3}).  The existence of these wave 
functions can be explored experimentally.

        A detailed examination of the band structure of BECs in periodic
 lattices requires a more intricate perturbative analysis than that discussed
 earlier in this work.  Previous authors have concentrated on numerical 
studies of band structure.\cite{band,space1,space2}  The approach we take, on
 the other hand, is to analyze the spatial resonance structure that arises 
from the nonlinear Mathieu equation obtained upon the application of a 
coherent structure ansatz to the cubic NLS.  We examine situations with null 
angular momentum ($c = 0$), but one observes similar behavior when $c \neq 0$
 when $R$ is away from the origin.  The analytical approach we employ was 
introduced by Zounes and Rand\cite{zounes} for $g < 0$ and $\mu > 0$ (see 
Figure \ref{repulse1}b), the technically easiest case to consider.  Their 
study of nonlinear Mathieu equations is directly applicable to BECs.  Our 
work is an extension of their work to the situation $g > 0$, $\mu > 0$ (see 
Figure \ref{repulse1}a), the second easiest case to consider.  We study this 
case in detail and also apply the results of Zounes and Rand to 
attractive BECs with a positive chemical potential.  We briefly discuss 
attractive BECs with a negative chemical potential (see 
Figure \ref{repulse1}c), the technically hardest case to consider.  Note 
that this paper does {\it not} explore the chaotic dynamics of BECs, which 
is an important open issue.\cite{gucken,wiggins,rand,675} 

Let $x_0 = -\pi/(2\kappa)$ and $V_1 (x) \equiv 0$ so that 
\begin{equation}
        V(x) = V_0\cos(\kappa x) \,. 
\end{equation}
When $c = 0$, the equations of motion (\ref{harmwiggle}), 
(\ref{dynam35}) for the amplitude of the coherent structure (\ref{maw2}) take
 the form
\begin{equation}
        R'' + \delta R + \alpha R^3 + \epsilon R \cos(\kappa x) = 0 \,, 
\label{dyn}
\end{equation}
where   
\begin{equation}
        \delta = \frac{2m\mu}{\hbar} \,, \quad
        \alpha = -\frac{2mg}{\hbar^2} \,, \quad
        \epsilon = -\frac{2m}{\hbar^2}V_0 \,.
\end{equation}
(Note that the perturbation parameter $\epsilon$ is not the same as the 
parameter $\varepsilon$ employed earlier.)
The parameters $\mu$, $V_0$, $\kappa$, and $a$ (and hence $g$) can all be 
adjusted experimentally.  When $\epsilon = 0$, solutions of (\ref{dyn}) can 
be written exactly in terms of elliptic 
functions:\cite{zounes,coppola,copthes,rand,lawden,watson}
\begin{equation}
        R = \sigma \rho \, \mbox{cn}(u,k) \,, 
\label{ell0}
\end{equation}
where
\begin{align}
        u &= u_1x + u_0 \,, \quad u_1^2 = \delta + \alpha \rho^2 \,, \notag \\
k^2 &= \frac{\alpha \rho^2}{2(\delta + \alpha \rho^2)} \,, \notag \\ 
u_1 &\geq 0, \hspace{.1 in} \rho \geq 0\, , \hspace{.1 in} k^2 \in 
\mathbb{R} \,, \hspace{.1 in} \sigma \in \{-1,1\} \,, \label{ellipse}
\end{align}
and $u_0$ is obtained from an initial condition (and can be set to $0$ 
without loss of generality).  We consider $u_1 \in \mathbb{R}$ in order to 
study periodic solutions.  One can use argument transformations to study
 solutions with complex $u_1$.  When $k^2 \in (1,\infty)$, one makes sense 
of the $\mbox{cn}$ function with a reciprocal modulus 
transformation.\cite{coppola,copthes}  When $k^2 < 0$, one employs a 
reciprocal complementary modulus transformation, which we discuss below.  

Equation (\ref{dyn}) can be integrated when $\epsilon = 0$ to yield the 
Hamiltonian
\begin{equation}
        \frac{1}{2}R'^2 + \frac{1}{2}\delta R^2 + \frac{1}{4}\alpha R^4 
= h\,, 
\end{equation}
with given energy $h$.  With (\ref{ellipse}), one computes
\begin{equation}
        h = \frac{1}{4}\rho^2(2\delta + \alpha\rho^2) 
= \frac{\delta^2}{\alpha}\frac{k^2 k'^2}{(1-2k^2)^2} \,,
\end{equation}
where $k'^2 := 1 - k^2$.  Earlier in this paper, we enumerated the different 
possibilities for the qualitative dynamics of (\ref{dyn}) in terms of 
the signs of $\mu$ and $g$ (and hence in terms of the signs of $\delta$ and 
$\alpha$).

\subsection{Repulsive BECs with a Positive Chemical Potential} \label{anal}

        We first consider in detail the case $g > 0$, $\mu > 0$, for 
which $\delta > 0$, $\alpha < 0$.  For notational convenience, we sometimes 
utilize $\alpha' := -\alpha$.  This analysis involves a considerable amount 
of elliptic-function manipulation, but we are rewarded in the end by a much 
more effective perturbation theory than can be obtained by employing 
trigonometric functions.

The center at $(0,0)$ satisfies $h = \rho^2 = k^2 = 0$.  The saddles at 
$(\pm\sqrt{\delta/\alpha'},0)$ and their adjoining separatrix satisfy 
\begin{equation}
        h = -\frac{\delta^2}{4\alpha} \,, \hspace{.1 in} 
\rho^2 = \frac{\delta}{|\alpha|} \,, \hspace{.1 in} k^2 = -\infty \,.
\end{equation}
The sign $\sigma = +1$ is used for the right saddle, and $\sigma = -1$ is 
used for the left one.  Within the separatrix, all orbits are periodic and 
the value of $\sigma$ is immaterial.

\subsubsection{Action-Angle Variable Description and Transformations}

For this choice of parameters, $k^2 \in [-\infty,0]$, so elliptic 
functions are defined through the reciprocal complementary modulus 
transformation,\cite{coppola,copthes} which relates the $(u,k)$ coordinate 
system to another coordinate system, which we denote $(w,k_2)$.  To tranform 
between these two coordinate systems, one uses the following relations:
\begin{align}
        \mbox{cn}(u,k) &= \mbox{cd}(w,k_2) \,, \notag \\ 
        \mbox{dn}(u,k) &= \mbox{nd}(w,k_2) \,, \notag \\ 
        \mbox{sn}(u,k) &= k_2' \mbox{sd}(w,k_2) \,, \notag \\ 
        k' = \frac{1}{k_2'} \,, \; u &= k_2'w \,, \; K = k_2'K_2 \,, 
\; E = \frac{1}{k_2'}E_2 \,. \label{rcmt}
\end{align}
Here, $K \equiv K(k)$ denotes the complete elliptic integral of the
 first kind, $E \equiv E(k)$ denotes the complete elliptic integral of the 
second kind, and items with the subscript `2' denote the analogous quantities
 in the $(w,k_2)$ coordinate system.\cite{watson,lawden,copthes}

We rescale (\ref{dyn}) using the coordinate transformation
\begin{equation}
        \chi = \sqrt{\delta}x \,, \qquad r = \sqrt{\frac{\delta}{\alpha'}}R  
\label{nondim}
\end{equation}
to obtain
\begin{equation}
        r'' + r - r^3 = 0 \label{dyn2}
\end{equation}
when $V(x) \equiv 0$.  (Note that in this analysis, the quantity $\chi$ does 
not represent the mean healing length.)  In terms of the original coordinates,
\begin{equation}
        R(x) = \sqrt{\frac{\delta}{\alpha'}}r(\sqrt{\delta}x) 
= \sqrt{\frac{\hbar\mu}{g}}r\left(\sqrt{\frac{2m\mu}{\hbar}}x\right) 
\,.
\end{equation}
The rescaling applied for other choices of $\delta$ and $\alpha$ differ 
slightly from that in (\ref{nondim}), so that the arguments of their 
associated square roots are positive.

The Hamiltonian corresponding to (\ref{dyn2}) is
\begin{equation}
        H_0(r,s) = \frac{1}{2}s^2 + \frac{1}{2}r^2 - \frac{1}{4}r^4 = h \,, 
\; h \in [0, 1/4] \,,  \label{sheep}
\end{equation}
where $s := r' = dr/d\chi$.  Additionally, 
$\rho^2 \in [0,1]$, $k_2^2 \in [0,1]$ (corresponding to 
$k^2 \in (-\infty,0]$ in the original coordinates), and
\begin{equation}
        k^2 = \frac{\rho^2}{2(\rho^2 - 1)} \,.
\end{equation}
With the initial condition $r(0) = \rho$, $s(0) = 0$, which implies that 
$u_0 = 0$, solutions to (\ref{dyn2}) are given by
\begin{align}
        r(\chi) &= \rho \, \mbox{cn}\left(\left[1-\rho^2\right]^{1/2}\chi,k\right) \,, \notag \\
        s(\chi) &= -\rho\left[1-\rho^2\right]^{1/2}\,\mbox{sn}\left(\left[1-\rho^2\right]^{1/2}\chi,k\right) \notag \\ &\qquad \times \, \mbox{dn}\left(\left[1-\rho^2\right]^{1/2}\chi,k\right) \,. \label{soly}
\end{align}

The period of a given periodic orbit $\Gamma$ is
\begin{equation}
        T(k) = \oint_\Gamma d\chi = \frac{4K(k)}{\sqrt{1-\rho^2}} \,,
\end{equation}
where $4K(k)$ is the period in $u$ of $\mbox{cn}(u,k)$.\cite{watson}  The 
frequency of this orbit is
\begin{equation}
        \Omega(k) = \frac{\pi\sqrt{1-\rho^2}}{2K(k)} \,.
\end{equation}

Let $\Gamma_h$ denote the periodic orbit with energy $h = H_0(r,s)$.  The 
area of phase space enclosed by this orbit is constant with respect to 
$\chi$, so one may define the action\cite{goldstein,rand,coppola,copthes}
\begin{equation}
        J := \frac{1}{2\pi}\oint_{\Gamma_h}sdr 
= \frac{1}{2\pi}\int_0^{T(k)}[s(\chi)]^2d\chi \,,
\end{equation}
which in this case can be evaluated exactly:
\begin{equation}
        J = \frac{4\sqrt{1-\rho^2}}{3\pi}\left[E(k) 
- \left(1 - \rho^2/2\right)K(k)\right] \,. 
\end{equation}
The associated angle\cite{goldstein,gucken,wiggins,lich} in the canonical 
transformation $(r,s) \longrightarrow (J,\Phi)$ is 
\begin{equation}
        \Phi := \Phi(0) + \Omega(k)\chi \,.
\end{equation}
The frequency $\Omega(k)$ monotonically decreases as $k^2$ goes from 
$-\infty$ to $0$ [that is, as one goes from the separatrix to the center at 
$(r,s) = (0,0)$].  With this transformation, equation (\ref{soly}) becomes
\begin{align}
        r(J,\Phi) &= \rho(J) \, \mbox{cn}\left(2K(k)\Phi/\pi,k\right) 
\,, \notag \\
        s(\chi) &= -\rho(J)\sqrt{1-\rho(J)^2}\,\mbox{sn}\left(2K(k)\Phi/\pi,k\right) \notag \\ &\qquad \times \, \mbox{dn}\left(2K(k)\Phi/\pi,k\right) \,, \label{soly2}
\end{align}
where $k = k(J)$.

After rescaling, the equations of motion for the forced system (\ref{dyn}) 
take the form
\begin{equation}
        r'' + r - r^3 + \frac{\epsilon}{\delta}\cos\left(\frac{\kappa}{\sqrt{\delta}}\chi\right)r = 0
\end{equation}
with the corresponding Hamiltonian
\begin{align}
        H(r,s,\chi) &= H_0(r,s) + \epsilon H_1(r,s,\chi) \notag \\
        &= \frac{1}{2}s^2 + \frac{1}{2}r^2 - \frac{1}{4}r^4 + 
\frac{\epsilon}{2\delta}r^2\cos\left(\frac{\kappa}{\sqrt{\delta}}\chi\right) 
\,.
\end{align}
In action-angle coordinates, this becomes
\begin{align}
        H(\Phi,J,\chi) &= h(J) + \epsilon h_1(\Phi,J,\chi) \notag \\
        &= \frac{1}{2}\rho(J)^2 - \frac{1}{4}\rho(J)^4 \label{hamaction} \\ &+ \frac{\epsilon}{2\delta}\rho(J)^2 \, \mbox{cn}^2\left(2K(k)\Phi/\pi,k\right)\cos\left(\frac{\kappa}{\sqrt{\delta}}\chi\right) \,.  \notag
\end{align}

One obtains a second action-angle pair $(\phi,j)$ using the canonical 
transformation $(\Phi,J) \longrightarrow (\phi,j)$ defined by the relations
\begin{equation}
        j(J) = \frac{1}{2}\rho(J)^2 \,, 
\quad \Phi(\phi,j) = \frac{\phi}{J'(j)} \,,
\end{equation}
where
\begin{align}
        k^2 &= \frac{j}{2j-1} \,, \notag \\
        J(j) &= \frac{2}{3}\sqrt{1-2j}\left[\tilde{E}(j) 
- (1-j)\tilde{K}(j)\right] \,, \notag \\
        \tilde{K}(j) &= \frac{2}{\pi}K[k(j)] \,, 
\quad \tilde{E}(j) = \frac{2}{\pi}E[k(j)] \,.
\end{align}
Additionally,
\begin{equation}
        J'(j) := \frac{dJ}{dj} = \sqrt{1-2j}\tilde{K}(j) 
= \frac{1-2j}{\Omega(j)} \,. \label{fre}
\end{equation}
Note that $J \sim j$ for small-amplitude motion.  Furthermore, 
$j = 0$ at the origin, and $j = 1/2$ on the separatrix.

The Hamiltonian (\ref{hamaction}) becomes
\begin{align}
        H(\phi,j,\chi) &= H_0(j) + \epsilon H_1(\phi,j,\chi)  \label{haha} \\
        &= j - j^2 + \frac{\epsilon}{\delta}j \, \mbox{cn}^2\left(\frac{\tilde{K}(j)}{J'(j)}\phi,k\right)\cos\left(\frac{\kappa}{\sqrt{\delta}}\chi\right) \,. \notag
\end{align}

Because we have used elliptic functions rather than trigonometric functions, 
all results are exact thus far.\cite{zounes}

\subsubsection{Perturbative Analysis} 

A subsequent $\mathcal{O}(\epsilon)$ analysis at this stage allows one to 
study $2m'\!:\!1$ subharmonic resonances for all $m' \in \mathbb{Z}$.  By 
contrast, had we undertaken this procedure with trigonometric functions 
(which would have entailed a perturbative approach from the beginning), an 
$\mathcal{O}(\epsilon^{m'})$ analysis would be required to study $2m'\!:\!1$ 
subharmonic resonances of (\ref{dyn}).  

The Fourier expansion of $\mbox{cn}$ is given by
\begin{equation}
        \mbox{cn}(u,k) = \frac{2\pi}{kK(k)}\sum_{n = 0}^\infty b_n(k) \cos\left[(2n+1)\frac{\pi u}{2K(k)}\right] \,,
\end{equation}
where the Fourier coefficients $b_n(k)$ are
\begin{equation}
        b_n(k) = \frac{1}{2}\mbox{sech}\left[\left(n + 1/2\right)\pi K'(k)/K(k)\right] \,, \label{cof1}
\end{equation}
and $K'(k) := K(\sqrt{1-k^2})$ denotes the complementary complete elliptic 
integral of the first kind.\cite{watson,stegun,zounes}  In the present 
situation,
\begin{equation}
        \mbox{cn}\left(\frac{\tilde{K}(j)}{J'(j)}\phi,k\right) 
= \sum_{n = 0}^\infty B_n(j)\cos\left[(2n+1)\frac{\phi}{J'(j)}\right] \,,
\end{equation}
where
\begin{equation}
        B_n(j) = \frac{4}{k(j)\tilde{K}(j)}b_n[k(j)] \,. \label{cof2}
\end{equation}
Consequently, 
\begin{equation}
        \mbox{cn}^2\left(\frac{\tilde{K}(j)}{J'(j)}\phi,k\right) 
= \mathcal{B}_0(j) 
+ \sum_{l = 1}^\infty \mathcal{B}_l\cos\left(\frac{2l\phi}{J'(j)}\right) \,,
\end{equation}
where the Fourier coefficients $\mathcal{B}_l(j)$ are obtained by 
convolving the previous Fourier coefficients (\ref{cof2}) with each 
other.\cite{zounes}

Before proceeding, it is important to discuss the computation of the 
coefficients $\mathcal{B}_l(j)$, which require some care.  Using the Elliptic
 Nome\cite{stegun,crc}
\begin{equation}
        q(k) := e^{-\pi K'(k)/K(k)} \,,
\end{equation}
the Fourier coefficient (\ref{cof1}) is expressed as
\begin{equation}
        b_n(k) = \frac{1}{q(k)^{n+1/2} + q(k)^{-(n + 1/2)}} \,.
\end{equation}
One then expands $\mathcal{B}_l(j)$ in Taylor series about $j = 0$.  In this 
computation, one finds that the coefficients of even powers of $j$ in 
$\mathcal{B}_l(j)$ are the same as when $g < 0$, $\mu > 0$ and that odd 
powers have the opposite sign.  This distinction lies at the root of the 
qualitatively different dynamics in the two cases, which we will discuss in 
Section \ref{contrast}.  Recall that their underlying integrable dynamics are 
depicted in Figure \ref{repulse1}. 

After the Fourier expansion, the perturbative term in the Hamiltonian 
(\ref{haha}) is
\begin{align}
        H_1(\phi,j,\chi) &= \frac{1}{\delta}\mathcal{B}_0(j)\cos\left(\frac{\kappa}{\sqrt{\delta}}\chi\right) \notag \\ &+ \frac{1}{2\delta}\sum_{l = 1}^\infty \mathcal{B}_l(j)\left[\cos\left(\frac{2l\phi}{J'(j)} + \frac{\kappa}{\sqrt{\delta}}\chi\right) \right. \notag \\  &\qquad + \left. \cos\left(\frac{2l\phi}{J'(j)} - \frac{\kappa}{\sqrt{\delta}}\chi\right)\right] \,. \label{haha2}
\end{align}

There are infinitely many (subharmonic) resonance 
bands,\cite{gucken,wiggins,rand} each of which corresponds to a single 
harmonic in the perturbation series (\ref{haha2}).  To isolate individual
 resonances, we apply a canonical, near-identity 
transformation\cite{zounes,rand,gucken,wiggins,goldstein} to the Hamiltonian 
$H = H_0 + \epsilon H_1$.  This transformation is given by
\begin{align}
        \phi &= Q + \epsilon \frac{\partial W_1}{\partial P} 
+ \mathcal{O}(\epsilon^2) \,, \notag \\
        j &= P - \epsilon \frac{\partial W_1}{\partial Q} 
+ \mathcal{O}(\epsilon^2) \,,
\end{align}
where the generating function $W_1$ is
\begin{align}
        W_1 &= \frac{P\mathcal{B}_0(P)}{\kappa\sqrt{\delta}}\sin\left(\frac{\kappa}{\sqrt{\delta}}\chi\right) \notag \\ &+ \frac{P}{2\sqrt{\delta}}\sum_{l=1\,,\,l \neq m'}^\infty \mathcal{B}_l(P)\left[\frac{\sin\left(\frac{2lQ}{J'(P)} + \frac{\kappa}{\sqrt{\delta}}\chi\right)}{\kappa + 2l\sqrt{\delta}\Omega(P)} \right. \notag \\ &\qquad \qquad + \left. \frac{\sin\left(\frac{2lQ}{J'(P)} - \frac{\kappa}{\sqrt{\delta}}\chi\right)}{\kappa - 2l\sqrt{\delta}\Omega(P)}\right] \,. \label{gf}
\end{align}
To obtain (\ref{gf}), one uses the fact [from (\ref{fre})] that 
$\Omega(P) = (1-2P)/J'(P)$.

The resulting Hamiltonian is
\begin{align}
        K(Q,P,\chi) &= K_0(P) + \epsilon K_1(Q,P,\chi) \,, \notag \\
        K_0(P) &= P - P^2 = H_0(P) \,, \notag \\
        K_1(Q,P,\chi) &= H_1(Q,P,\chi) + \{H_0,W_1\} 
- \frac{\partial W_1}{\partial \chi} \,,
\end{align}
where $\{A_1,A_2\}$ denotes the Poisson bracket of $A_1$ and $A_2$.  For the 
present choice of $W_1$, one obtains the resonance Hamiltonian
\begin{align}
        K(Q,P,\chi;m') &= P - P^2 \notag \\
&\hspace{-.5 in} + \frac{\epsilon}{2\delta}P\mathcal{B}_{m'}(P)\cos\left(\frac{2m'Q}{J'(P)} 
- \frac{\kappa}{\sqrt{\delta}}\chi\right) + \mathcal{O}(\epsilon^2) \,. 
\label{kam1}
\end{align}
The choice of the generating function (\ref{gf}) eliminates all resonances 
from the Hamiltonian $K$ except the $2m'\!:\!1$ resonance.  In focusing on a 
single resonance band in phase space, one restricts $P$ to a neighborhood
 of $P_{m'}$, which denotes the location of the $m'$th resonant torus 
(associated with periodic orbits in $2m'\!:\!1$ spatial resonance with the
 periodic lattice).

\subsubsection{Resonance Relations}

        Resonant frequencies arise when the denominators of the terms in
$W_1$ vanish,\cite{zounes,rand,gucken} which yields the equation
\begin{equation}
        \frac{\kappa}{\sqrt{\delta}} = \pm 2m'\Omega(P_m) \label{relate}
\end{equation}
for the resonance of order $2m'\!:\!1$.  As $\Omega \leq 1$ is a decreasing 
function of $P \in [0,1/2)$, the resonance band associated with $2m'\!:\!1$ 
subharmonic spatial resonances is present when
\begin{equation}
        \frac{\kappa}{\sqrt{\delta}} \leq 2m' \,. \label{onset}
\end{equation}
For example, when $\kappa = 2.5$ and $\delta = 1$, there are resonances of 
order $4\!:\!1$, $6\!:\!1$, $8\!:\!1$, {\it etc}, but there are no resonances
 or order $2\!:\!1$.  When $\kappa = 5$ and $\delta = 1$, there are 
resonances of order $6\!:\!1$, $8\!:\!1$, $10\!:\!1$, {\it etc}, but there 
are no resonances of order $2\!:\!1$ or $4\!:\!1$.  In terms of the original 
parameters, the condition (\ref{onset}) describing the onset of $2m'\!:\!1$ 
resonance bands takes the form
\begin{equation}
        \kappa \leq 2m'\sqrt{\frac{2m\mu}{\hbar}} \,. \label{onset2}
\end{equation}

If the lattice $V(x)$ has a smaller wave number (larger periodicity), then 
the chemical potential $\mu$ 
must be smaller for a given resonance to occur.  As $\kappa$ is decreased for
 a fixed $\mu$ (i.e., $\delta$) or as $\mu$ is increased for 
a given lattice size $\kappa$, resonance bands of lower order emerge from the
 origin and propagate in phase space.  Consequently, a sufficiently high 
order resonance is always present in (\ref{dyn}), but a given number of 
low-order ones may not be.  Lower-order resonances occupy larger regions of 
phase space, so (\ref{onset}) also indicates the volume of phase space 
affected by spatial resonances.  We will illustrate this in more detail in 
Section \ref{numsect} with numerical simulations.

\subsubsection{Analytical Description of Resonance Bands}

To further examine the resonance structure of (\ref{dyn}), we make 
(\ref{kam1}) autonomous via another canonical change of 
coordinates.\cite{zounes,rand}  Toward this end, we define the generating 
function
\begin{equation}
        F(Q,Y,\chi;m') = QY - \frac{\kappa}{2m'\sqrt{\delta}}J(Y)\chi \,,
\end{equation}
which yields
\begin{align}
        P &= \frac{\partial F}{\partial Q}(Q,Y,\chi) = Y \,, \notag \\
        \xi &= \frac{\partial F}{\partial Y}(Q,Y,\chi) \notag \\ &\qquad = Q 
- \frac{\kappa}{2m'\sqrt{\delta}}J'(Y)\chi = Q 
- \frac{\kappa}{2m'\sqrt{\delta}}J'(P)\chi \,.
\end{align}
(Note that in this analysis, $\xi$ does not represent stretched space, as it 
did in our multiple scale expansion.)  With this final transformation, the 
resonance Hamiltonian (\ref{kam1}) becomes
\begin{align}
        K_{m'}(\xi,Y) &= K(Q,P,\chi;m') 
+ \frac{\partial F}{\partial \chi}(Q,Y,\chi) \notag \\ 
&= Y - Y^2 - \frac{\kappa}{2m'\sqrt{\delta}}J(Y) \notag \\ &\qquad + \frac{\epsilon}{2\delta}Y\mathcal{B}_{m'}(Y)\cos\left(\frac{2m'\xi}{J'(Y)}\right) \,, \label{ram}
\end{align}
which is integrable in the $(Y,\xi)$ coordinate system.   In $(R,S)$-space, 
level curves of $K_{m'}$ correspond to invariant curves of Poincar\'e 
sections of (\ref{dyn}), which are defined by strobing the system when the 
spatial variable takes the values $x_n = 2n\pi/\kappa$.  

We now provide an analytical description of the resonance bands under 
discussion.  In particular, we compute the locations and type of equilibria 
and width of resonance bands as functions of the parameters $\delta$, 
$\epsilon$, and $\kappa$, and hence of $\mu$, $V_0$, and $\kappa$.  Such 
bands emerge from the action $P = Y = Y_{m'}$, which designates the location 
of the $m'$th resonance torus in phase space and is determined by the 
resonance relation (\ref{relate}):
\begin{equation}
        \frac{\kappa}{\sqrt{\delta}} = 2m'\Omega(Y_{m'}) \,. \label{resrel}
\end{equation}
This resonance band is associated with periodic orbits in $2m'\!:\!1$ spatial
 resonance with the periodic lattice.  The resonance torus is filled with 
degenerate periodic orbits that split\cite{gucken,wiggins} into $2m'$ saddles
 and $2m'$ centers when a perturbation is introduced.

From (\ref{ram}), one obtains Hamilton's equations
\begin{align}
        Y' &= -\frac{\partial K_{m'}}{\partial \xi} = \frac{\epsilon Y \mathcal{B}_{m'}(Y)}{J'(Y)}\sin\left(\frac{2m'\xi}{J'(Y)}\right), \notag \\
        \xi' &= \frac{\partial K_{m'}}{\partial Y} = 1 - 2Y - \frac{\kappa}{2m'\sqrt{\delta}}J'(Y) \notag \\ &\qquad + \frac{\epsilon}{2\delta}\Bigg[\left(Y\mathcal{B}_{m'}(Y)\right)'\cos\left(\frac{2m'\xi}{J'(Y)}\right)  \notag \\ &\qquad +  2m'\xi\frac{J''(Y)}{\left[J'(Y)\right]^2}Y\mathcal{B}_{m'}(Y)\sin\left(\frac{2m'\xi}{J'(Y)}\right)\Bigg] \,.
\end{align}
Equilibria satisfy either $Y = 0$ or
\begin{equation}
        \sin\left(\frac{2m'\xi}{J'(Y)}\right)  = 0\,.  \label{sineq}
\end{equation}
They also satisfy
\begin{equation}
        \xi' = 0 = 1 - 2Y - \frac{\kappa}{2m'\sqrt{\delta}}J'(Y) 
\pm \frac{\epsilon}{2\delta}\left[Y\mathcal{B}_{m'}(Y)\right]' \,, 
\label{cond}
\end{equation}
where the sign $\pm$ in (\ref{cond}) arises from
\begin{equation}
        \cos\left(\frac{2m'\xi}{J'(Y)}\right)  = 0\,.
\end{equation}
Using $J'(Y) = \sqrt{1-2Y}\tilde{K}(Y)$, equation (\ref{cond}) is written
\begin{equation}
        1 - 2Y - \frac{\kappa}{2m'\sqrt{\delta}}\sqrt{1-2Y}\tilde{K}(Y) 
\pm \frac{\epsilon}{2\delta}\left[Y\mathcal{B}_{m'}(Y)\right]' = 0 \,. 
\label{6point5}
\end{equation}

When $g < 0$ and $\mu > 0$, the $+$ case yields a saddle and the $-$ 
case yields a center.  In the present situation $(g > 0,\, \mu > 0)$, this
 holds for odd $m'$.  When $m'$ is even, $-$ is a saddle and $+$ is a center.

At equilibria, the action $Y$ takes the value
\begin{equation}
        Y_e = Y_{m'} + \epsilon \Delta Y + \mathcal{O}(\epsilon^2) 
= Y_{m'} \pm \mathcal{O}(\epsilon) \,, \label{ansa}
\end{equation}
with the signs as in (\ref{6point5}).  However, note that 
$Y_c > Y_{m'} > Y_s$, just as for $g < 0$.  One inserts (\ref{ansa}) into 
(\ref{6point5}) and expands the result in a power series.  At order 
$\mathcal{O}(\epsilon^0) = \mathcal{O}(1)$, this reproduces the resonance 
relation (\ref{resrel}).  At order $\mathcal{O}(\epsilon)$, one obtains
\begin{equation}
        \Delta Y = \mp\frac{\epsilon}{2\delta}\left[\frac{\mathcal{B}_{m'}(Y_{m'}) + Y_{m'}\mathcal{B}_{m'}'(Y_{m'})}{\Omega(Y_{m'})\sqrt{1-2Y_{m'}}\tilde{K}'(Y_{m'}) - 1}\right] \,, \label{deldel}
\end{equation}
where saddles $Y_s$ use the $+$ sign and centers $Y_c$ use the $-$ sign when 
$m'$ is even, and the opposite is true when $m'$ is odd.  When $m'$ is even, 
$\Delta Y > 0$, but $\Delta Y < 0$ when $m'$ is odd.  Additionally, $Y_c$ is 
always larger than $Y_s$ (for both signs of $\epsilon$).

Resonance bands occupy a finite region of phase space bounded by a 
pendulum-like separatrix.  When a perturbation is introduced, trajectories 
outside the separatrix behave almost as they would in the absence of a 
perturbation, so it is important to estimate the width of resonance bands, 
which emerge at action values satisfying the resonance relation 
(\ref{resrel}).  Because of the direction of the inequality in 
(\ref{onset2}), this is more of a condition for non-existence of given 
resonances.  [See the discussion following equation (\ref{onset}).]  For a 
given set of parameters, there will always be resonances of sufficiently high
 order (i.e., for a sufficiently large $m'$).  However, as we illustrate 
numerically below, there are parameter regions in which no $2\!:\!1$ 
resonances exist, regions in which no $2\!:\!1$ or $4\!:\!1$ resonances 
exist, {\it etc}.  This behavior contrasts markedly with that observed when 
$g < 0$.\cite{zounes}  In that situation, there exist parameter regions in 
which only  $2\!:\!1$ resonances exist, regions in which only  $2\!:\!1$ and 
 $4\!:\!1$ resonances exist, {\it etc}.

We now show that the width of a resonance band is
\begin{equation}
        \mathcal{O}\left(\sqrt{\frac{Y_{m'}\mathcal{B}_{m'}(Y_{m'})|\epsilon|}
{\delta}}\right)
\end{equation}
for perturbations of size $\epsilon = -2mV_0/\hbar^2$.

The separatrix of interest passes through the saddle point $Y_s$, and the 
maximum extent of the resonance band occurs at the same phase $\xi$ as the 
associated center, so
\begin{align}
        &K_{m'}\left(\cos\left(\frac{2m'\xi}{J'(Y_s)}\right) = -1,Y = Y_s\right) 
\notag \\ 
        &\qquad = K_{m'}\left(\cos\left(\frac{2m'\xi}{J'(Y)}\right) = +1,Y\right)
\end{align}
when $m'$ is odd and
\begin{align}
        &K_{m'}\left(\cos\left(\frac{2m'\xi}{J'(Y_s)}\right) = +1,Y = Y_s\right) 
\notag \\
        &\qquad = K_{m'}\left(\cos\left(\frac{2m'\xi}{J'(Y)}\right) = -1,Y\right)
\end{align}
when $m'$ is even.  This implies that
\begin{align}
        (Y - Y_s) &- (Y - Y_s)^2 
- \frac{\kappa}{2m'\sqrt{\delta}}(J(Y) - J(Y_s) \notag \\
&\pm \frac{\epsilon}{2\delta}(Y\mathcal{B}_{m'}(Y) + Y_s\mathcal{B}_{m'}(Y_s)) 
= 0 \,, \label{width}
\end{align}
where the $+$ sign holds for odd $m'$ and the $-$ sign holds for even $m'$.  
(Only the $+$ case needs to be considered when $g < 0$ and $\mu > 0$.)

Solutions $Y_*$ of (\ref{width}) are perturbations to $Y_{m'}$ of the form
\begin{equation}
        Y_* = Y_{m'} + W\epsilon^\gamma + \mathcal{O}(\epsilon^{2\gamma}) 
\label{gammy}
\end{equation}
for an appropriate choice of $\gamma$, to be determined by a self-consistency
 argument.  (When $\epsilon < 0$, one writes (\ref{gammy}) with 
$(-\epsilon)^\gamma$ instead.  Everything stated here is otherwise the same 
in that situation.)  In this analysis, one uses the fact that 
$Y_s = Y_{m'} \pm \epsilon (\Delta Y) + \mathcal{O}(\epsilon^2)$, where the 
$+$ sign is for odd $m'$ and the $-$ sign is for even $m'$.  

To find $W$ and $\gamma$, we insert $Y_*$ and $Y_s$ into (\ref{width}) and 
expand the resulting expression in a power series about $\epsilon = 0$.  
During this process, one obtains
\begin{align}
        Y_*^2 &= Y_{m'}^2 + 2Y_{m'}W\epsilon^\gamma + W^2\epsilon^{2\gamma}\,,
 \notag \\
        J(Y_*) &= J(Y_{m'}) + \epsilon^\gamma W J'(Y_{m'}) \notag \\
&\qquad + \epsilon^{2\gamma}W^2 J''(Y_{m'}) + \mathcal{O}(\epsilon^{3\gamma}) 
\,,
\end{align}
which shows that that the only suitable value of $\gamma$ is $1/2$.  Equating
 terms of order $\mathcal{O}(1)$ yields no new information.  Equating 
terms of order $\mathcal{O}(\epsilon^{1/2})$ yields the resonance relation 
(\ref{resrel}).  Equating terms of order $\mathcal{O}(\epsilon)$ shows 
that
\begin{equation}
        W = \left[ \pm\frac{Y_{m'}\mathcal{B}_{m'}(Y_{m'})}{\delta\left[1 + \frac{\kappa}{2m'\sqrt{\delta}}J''(Y_{m'})\right]}\right]^{1/2} \,, \label{dubya}
\end{equation}
where the $+$ sign occurs for odd $m'$ and the $-$ sign occurs for 
even $m'$.  Therefore, the miminal action of the resonance band is
\begin{equation}
        Y_{min} = Y_{m'} - \sqrt{\epsilon}W + \mathcal{O}(\epsilon) \,, 
\label{ymin}
\end{equation}
and the maximal action is 
\begin{equation}
        Y_{max} = Y_{m'} + \sqrt{\epsilon}W + \mathcal{O}(\epsilon) \,. 
\label{ymax}
\end{equation}
The width of the resonance band is $Y_{max} - Y_{min} = 2\sqrt{\epsilon}W 
+ \mathcal{O}(\epsilon)$.

In Section \ref{numsect}, we compare these analytical results with numerical 
simulations.

\subsection{Attractive BECs with a Positive Chemical Potential} 
\label{contrast}

        Zounes and Rand\cite{zounes} considered (\ref{dyn}) when $\delta > 0$ 
and $\alpha > 0$ (in other words, $\mu > 0$ and $g < 0$), which is 
depicted in Figure \ref{repulse1}b.  They did not consider the application 
of their analysis to Bose-Einstein condensates, so we presently interpret 
their results in this new light and compare it to our analysis of the 
repulsive case.  When $\delta > 0$, $\alpha > 0$, the phase space of the 
integrable problem contains no separatrix, and the entire space is foliated 
by periodic orbits.  (See Table \ref{tab2}.)  This choice of parameters also 
leads to the simplest application of the perturbation technique described in 
Section \ref{anal}.  In this case, $k \in [0,1]$, so one need not apply a 
modulus transformation in the elliptic function solution.  One may also set 
$\sigma = 1$.

We refer the reader to Zounes and Rand\cite{zounes} for details.  Here, we 
highlight a few results that we wish to contrast directly.  When $g < 0$ 
and $\mu > 0$, the resonance relation one obtains is
\begin{equation}
        \frac{\kappa}{\sqrt{\delta}} = 2m'\Omega_a(P_{m'}) \,,
\end{equation}
where the frequency $\Omega_a(P)$ has a similar form to that of $\Omega$ 
described above.  In this situation, $\Omega_a(P) \geq 1$, so subharmonic 
periodic orbits are present when
\begin{equation}
        \frac{\kappa}{\sqrt{\delta}} \geq 2m' \,, \label{reszr}
\end{equation}
which is the reverse inequality as that derived in the repulsive case.  
Hence, there exist regimes in which {\it only} $2\!:\!1$ resonances are 
present, {\it only} $2\!:\!1$ and $4\!:\!1$ resonances are present, 
{\it etc}.  In terms of BEC parameters, the condition (\ref{reszr}) describing
 the onset of $2m'\!:\!1$ resonance bands takes the form
\begin{equation}
        \kappa \geq 2m'\sqrt{\frac{2m\mu}{\hbar}} \,. \label{onset3}
\end{equation}
Because the inequalities in (\ref{onset2}) and (\ref{onset3}) are oppositely 
directed, adjustments to $\kappa$ and $\mu$ have the opposite effect in 
these two cases.
 
Additionally, in this case there is no alternating of signs in the location 
of saddles and centers in resonance bands, as there is when $g > 0$ and 
$\mu > 0$.  Because the attractive case with a positive chemical potential 
is simpler than the one we studied, Zounes and Rand\cite{zounes} were able to
 obtain better predictions describing the location of saddles and centers and
 the width of resonance bands from a perturbation analysis like that 
discussed in Section \ref{anal}.

\subsection{Attractive BECs with a Negative Chemical Potential}

        The most difficult case to consider is that of attractive BECs with a 
negative chemical potential.  In (\ref{dyn}), $\alpha > 0$ and $\delta < 0$ 
(i.e., $\mu < 0$), so the integrable dynamics exhibit two homoclinic 
orbits.  (See Figure \ref{repulse1}c.)  The perturbative approach used in this
 paper must be applied separately inside and outside the separatrix.  Orbits 
inside the separatrix satisfy $h < 0$, those on the separatrix satisfy 
$h = 0$, and those outside the separatrix satisfy $h > 0$.  

Inside the separatrix, $k \in (1,\infty)$, so one must apply the reciprocal 
modulus transformation to the arguments of the elliptic functions 
(\ref{ell0}), (\ref{ellipse}).  The sign of $\sigma$ determines whether one 
is considering perturbations of periodic orbits in the right half or left 
half of the phase plane.  To utilize our perturbative analysis outside the 
separatrix, one must expand elliptic functions and elliptic integrals in 
power series about infinity, where $k = 0$.  This requires delicate 
numerical computations of Laurent series coefficients.  

In principle, one can overcome the increased technical challenges present 
in this third case (which is also of interest) and apply the same analysis as
 in Section \ref{anal}, but the lengthy calculations involved would entail a 
separate publication.

\section{Numerical Simulations} \label{numsect}

To compare the analytical results in Section \ref{subharm} with numerical 
simulations, we utilize $(R,S)$ coordinates with $m = 1/2$ and $\hbar = 1$.  
To lowest order in $\epsilon$, the change of variables 
$Y \longrightarrow P \longrightarrow j$ is a near-identity transformation, so
 $Y = j + \mathcal{O}(\epsilon)$.  Recall from (\ref{sheep}) that
\begin{align}
        j &= \frac{1}{2}\rho^2, \notag \\
        H_0 &= h(j) = j - j^2 = \frac{1}{2}s^2 + \frac{1}{2}r^2 
- \frac{1}{4}r^4 \,,
\end{align}
where $s = \partial r/\partial \chi$.  For this comparison, we let 
$\alpha' = 1$ and vary $\kappa$, $\delta \equiv \mu$, and 
$\epsilon \equiv -2mV_0/\hbar^2$.  Recall additionally from (\ref{nondim}) that
\begin{equation}
        r = \sqrt{\frac{g}{\hbar\mu}}R 
= \sqrt{\frac{\alpha'}{\delta}}R \,, \quad 
        s = \frac{1}{\mu}\sqrt{\frac{g}{2m}}R' 
= \frac{\sqrt{\alpha'}}{\delta}R' \,.
\end{equation}

\subsection{Methodology}

Before discussing our results, we briefly overview our comparison procedure.

The ``exact'' locations of saddles and centers and sizes of resonance bands 
were determined using direct numerical simulations of Poincar\'e sections of
 (\ref{dyn}).  The surface of section we employed satisfies 
$x_n = 2n\pi/\kappa$ ($n \in \mathbb{Z}$), which consists of integer multiples 
of the periodicity of the sinusoidal forcing in (\ref{dyn}).  In our 
simulations, the variable $\kappa x$ is periodic, so the surface of section 
is defined by the condition $\kappa x = 0$.  We used this framework to find 
saddles, centers, and resonance band sizes (i.e., separatrix widths) 
empirically.

To obtain our predictions, we employed the resonance Hamiltonian 
(\ref{ram}), whose level curves correspond to invariant curves of Poincar\'e 
sections.  As each trajectory yields a level set of this Hamiltonian, 
we solved $K_m' = \mbox{constant}$ numerically at appropriate energy values to 
obtain predictions for the locations of saddles and centers and the size of 
resonance bands; these latter quantities are determined from the widths of 
separatrices in (\ref{ram}).  For these computations, we expanded elliptic 
functions and elliptic integrals in Taylor series and subsequently transformed 
these results to $(R,S)$-space to compare these calculations with our 
empirical ones.  We also predicted the locations of saddles and centers 
(\ref{ansa},\ref{deldel}) and the size of resonances bands 
(\ref{dubya},\ref{ymin},\ref{ymax}) using the predictions obtained from 
further perturbation expansions.  We again tranformed back to $(R,S)$ space to 
compare this second set of predictions with our empirical results.

\subsection{Primary Resonances}

Our comparison between theory and numerics for primary resonances is 
summarized in Tables \ref{primary1} and \ref{primary2}.

\begin{figure}
                \centerline{
                (a)
                \includegraphics[width=0.40\textwidth, height=0.50\textwidth]
{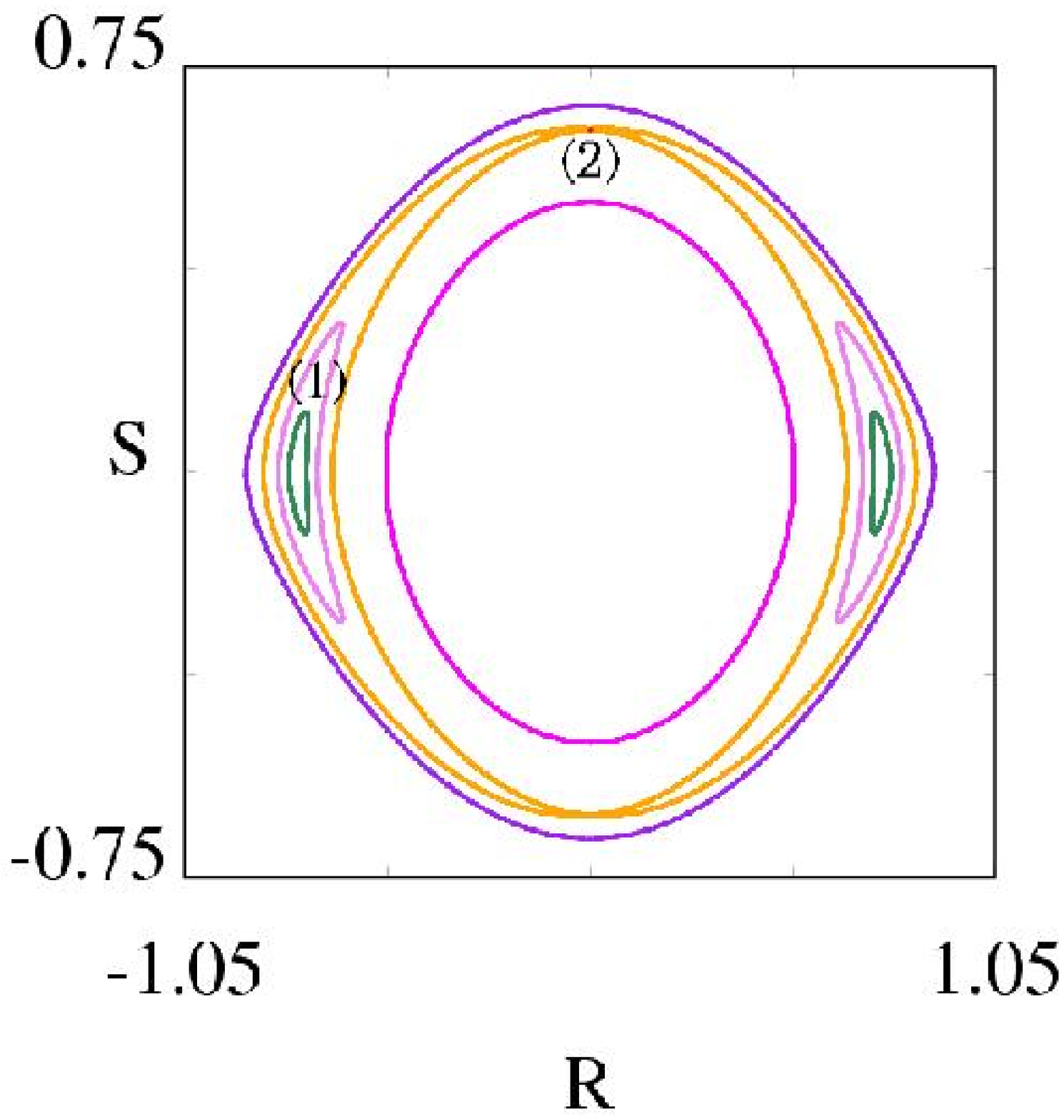}
                \hspace{1 cm}
                (b)
                \includegraphics[width=0.40\textwidth, height=0.50\textwidth]
{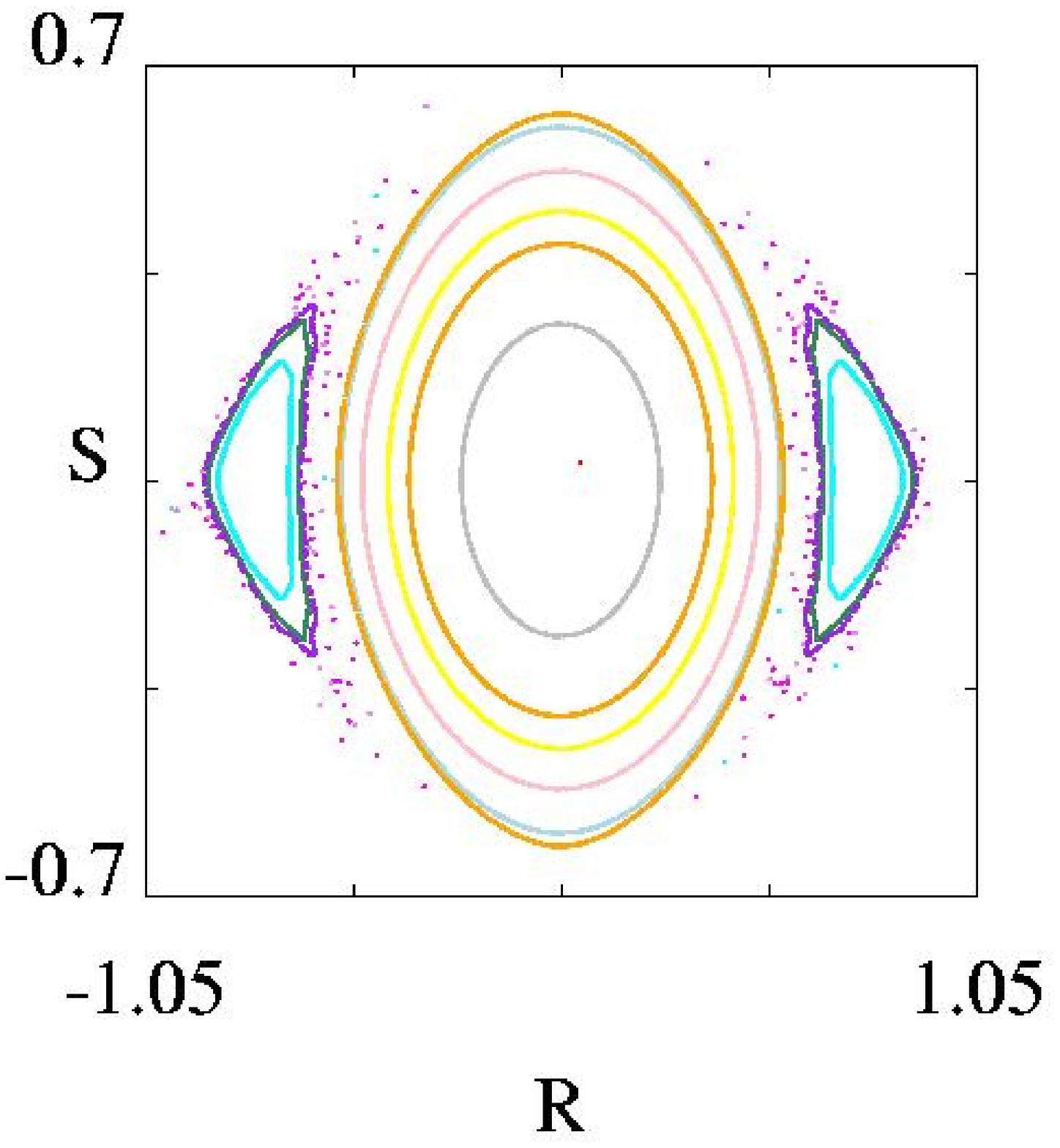}}
\vspace{.3 in}

                \centerline{
                (c)\includegraphics[width=0.40\textwidth, height=0.50\textwidth]
{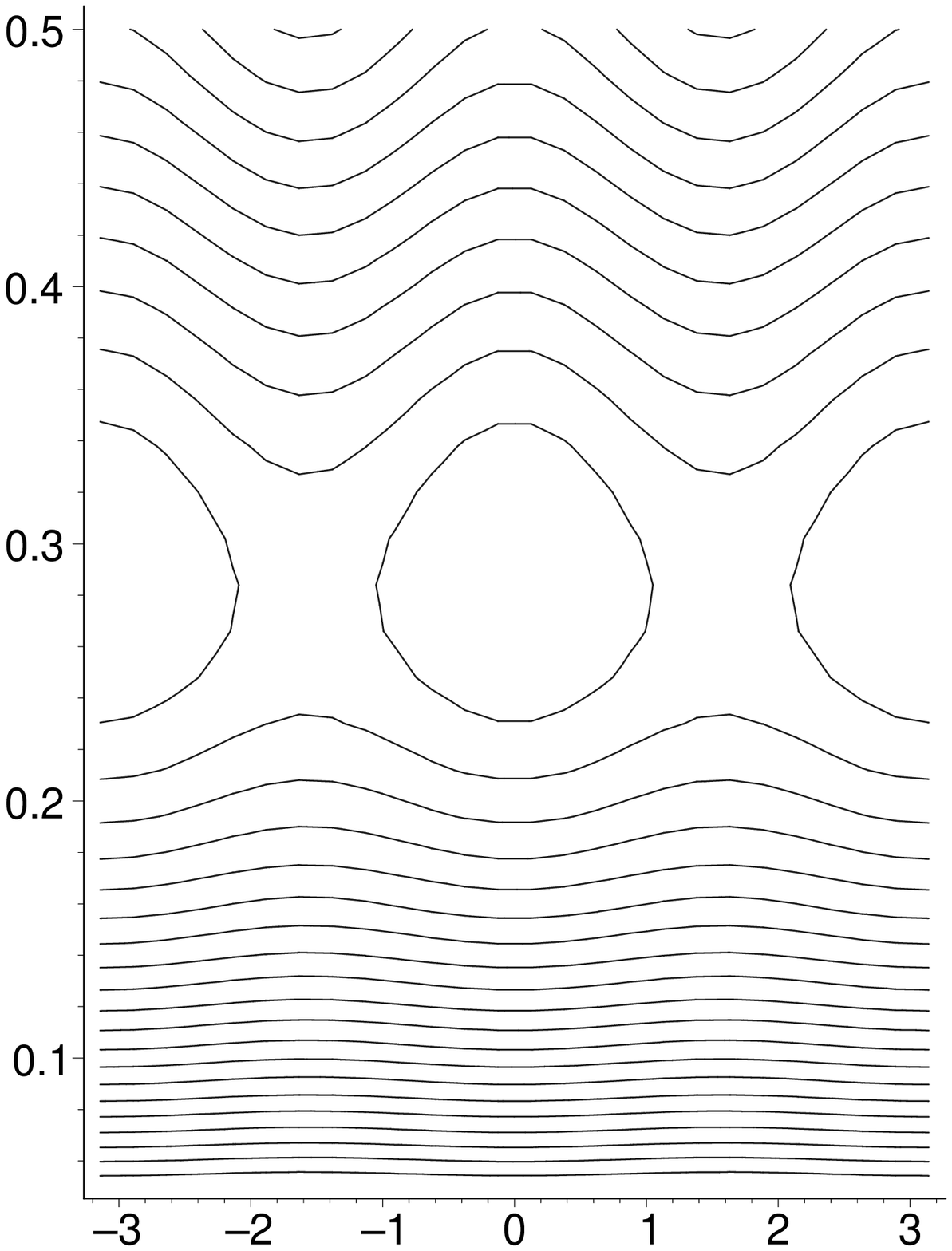}
                \hspace{1 cm}
                (d)
                \includegraphics[width=0.40\textwidth, height=0.50\textwidth]
{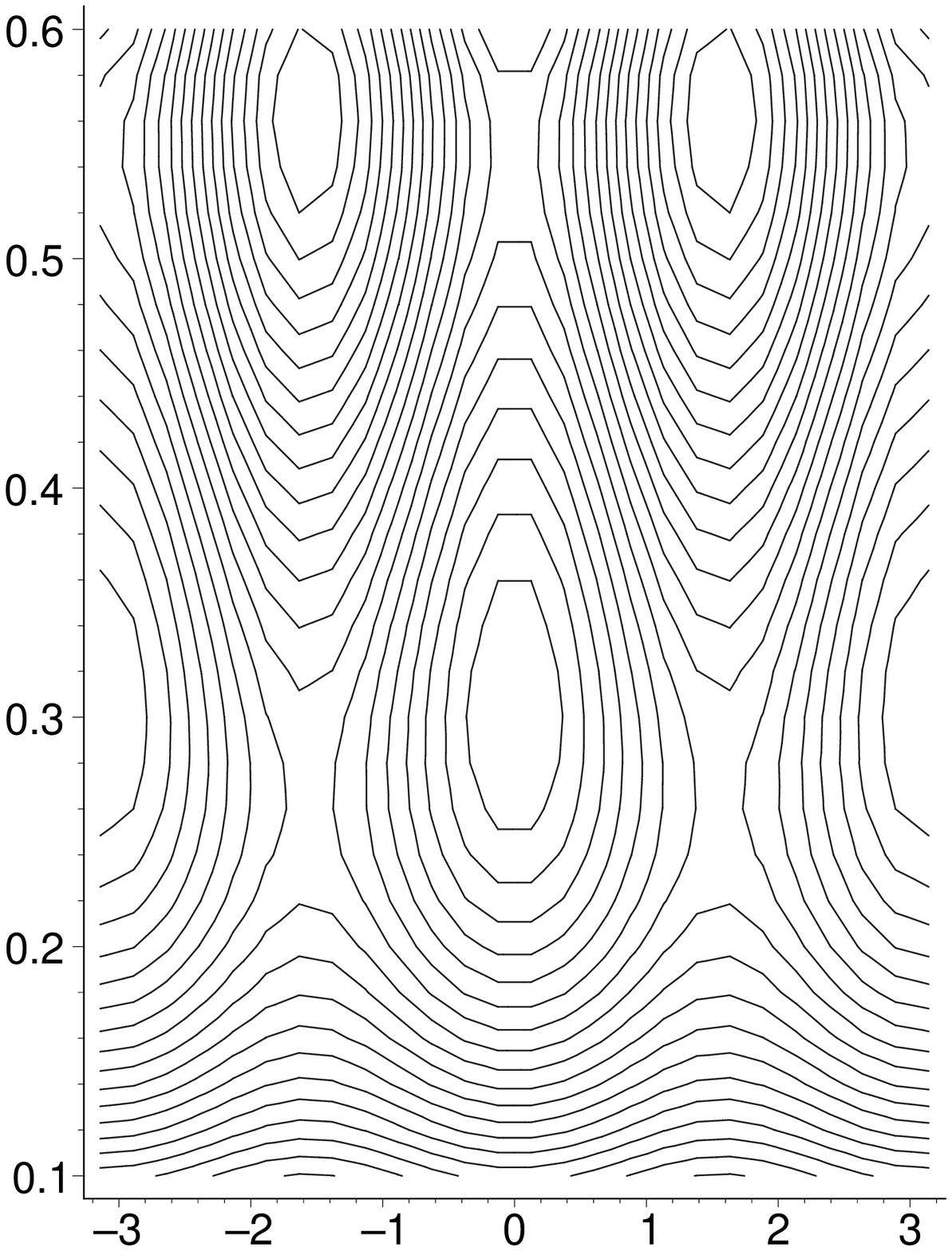}
                } 


                \caption{Poincar\'e sections (a), (b) and resonance 
Hamiltonians $K_1$ (c), (d) for $\kappa = 1.5$ and $\delta = 1$. (a)  
Poincar\'e section for $\epsilon = 0.01$.  The $2\!:\!1$ resonances are 
displayed, as indicated by the numbered trajectories.  (b) $\epsilon = 0.05$.
  (c) Resonance Hamiltonian for $\epsilon = 0.01$ with vertical axis in 
units of action $Y$ and horizontal axis in units of $\xi/J'(Y)$.  
(d) $\epsilon = 0.05$.} \label{res1}

\end{figure}

Consider first $\kappa = 1.5$ and $\delta = 1$.  Poincar\'e sections and 
level sets of the resonance Hamiltonian $K_1$ [in units of $\xi/J'(Y)$] are 
depicted in Figure \ref{res1}.  The results of our comparison between 
perturbation theory and numerical simulations are summarized in Table 
\ref{primary1}.

\begin{table}[t] 
\begin{tabular}{l r}
\hspace{4.8 cm} $\epsilon = 0.01$ \hspace{3.2 cm} & \hspace{-.75 cm} 
$\epsilon = 0.05$ \\  
\end{tabular}
\\
\centerline{
\begin{tabular}{|c||c|c||c|r|} \hline
Quantity & Perturbative & Numerical & Perturbative & Numerical \\ \hline
$Y_1$ & $0.28133$ & $\bullet$   & $0.28133$   & $\bullet$ \\
$Y_s$ & $0.27949$ & $\bullet$   & $0.27213$   & $\bullet$ \\
$R_s$ & $\pm 0.7477$ & $\pm 0.66$ & $\pm 0.73775$  & $\pm (0.56-0.66)$ \\
$Y_c$ & $0.28317$   & $\bullet$   & $0.29053$ & $\bullet$ \\
$R_c$ & $\pm 0.7526$ & $\pm 0.757$   & $\pm 0.76227$   & $\pm 0.774$ \\
$Y_{min}$ & $0.16175$ & $\bullet$   & ---   & $\bullet$ \\
$Y_{max}$ & $0.40091$   & $\bullet$   & --- & $\bullet$ \\
$R_{in}$ & $\pm 0.56877$ & $\pm 0.66$   & ---   & --- \\
$R_{out}$ & $\pm 0.89545$ & $\pm 0.85$   & ---   & --- \\
$Y_{min,2}$ & $0.22203$ & $\bullet$   & ---   & $\bullet$ \\
$Y_{max,2}$ & $0.35807$ & $\bullet$   & ---   & $\bullet$ \\
$R_{in,2}$ & $\pm 0.66638$ & $\pm 0.66$   & ---   & --- \\ 
$R_{out,2}$ & $\pm 0.84626$ & $\pm 0.85$   & ---   & --- \\ \hline
\end{tabular}}
\caption{Comparison of perturbation theory and numerics for $2\!:\!1$ 
resonances ($\kappa = 1.5$, $\delta = 1$).  In this Table, $Y_1$ is the 
action value of the primary resonance, $Y_s$ is the location of its 
nearby saddle, $(R_s,0)$ is its location in 
$(R,S)$-coordinates, $Y_c$ is the location of the nearby center, $(R_c,0)$ is 
its location in $(R,S)$-space, $Y_{min}$ is the minimum 
action value of the separatrix determined from (\ref{ymin}), $Y_{max}$ is the
 maximum determined from (\ref{ymax}), $R_{in}$ is where the 
inner separatrix crosses the $R$-axis, $R_{out}$ is where the outer 
separatrix crosses the $R$-axis, $Y_{min,2}$ and $Y_{max,2}$ are the minimum 
and maximum actions obtained by solving (\ref{width}) numerically, and 
$R_{in,2}$ and $R_{out,2}$ are their corresponding predictions of where the 
inner and outer separatrices cross the $R$-axis.  The symbol $\bullet$ means 
a calculation is not applicable and --- means it was not computed.} 
\label{primary1}
\end{table}

We do relatively well in locating saddles and extremely well in locating 
centers.  This is especially significant in light of the fact that many 
canonical transformations were required to obtain our analytical predictions.
  Although the requisite calculations are complicated, we are rewarded by 
excellent qualitative agreement and good (and sometimes excellent) 
quantitative agreement.  For $\epsilon = 0.05$, the numerical resolution of 
the location of the saddles was problematic, so a direct comparison is 
necessarily less accurate.  As a result, a range of values is sometimes 
indicated for the numerically determined location of saddles.  Such 
difficulties with direct numerical simulation emphasize the importance of 
using qualitative analytical methods to study the features of resonance 
bands. 

Our comparisons between perturbation theory and numerical simulations for 
$\kappa = 0.75$, $\delta = 0.2$ are summarized in Table \ref{primary2}.

If desired, one can improve these quantitative predictions by including 
higher-order contributions in the perturbation expansions.

\begin{table}[t] 
\begin{tabular}{l r}
$\hspace{4.9 cm} \epsilon = 0.01$ \hspace{3.35 cm} & \hspace{-.25 cm} 
$\epsilon = 0.05$ \\ 
\end{tabular}
\\
\centerline{
\begin{tabular}{|c||c|c||c|r|} \hline
Quantity & Perturbative & Numerical & Perturbative & Numerical \\ \hline
$Y_1$ & $0.19443$ & $\bullet$   & $0.19443$   & $\bullet$ \\
$Y_s$ & $0.18499$ & $\bullet$   & $0.14718$   & $\bullet$ \\
$R_s$ & $\pm 27202 $ & $\pm (0.19-0.20)$ & $\pm 0.24264$  & --- \\
$Y_c$ & $0.20389$   & $\bullet$   & $0.24169$ & $\bullet$ \\
$R_c$ & $\pm 0.28558$ & $\pm 0.2908$   & $\pm 0.31093$   & $\pm 0.335$ \\
$Y_{min}$ & $0.07485$ & $\bullet$   & ---   & $\bullet$ \\
$Y_{max}$ & $0.31402$   & $\bullet$   & --- & $\bullet$ \\
$R_{in}$ & $\pm 0.17303$ & $\pm 0.19$   & ---   & --- \\
$R_{out}$ & $\pm 0.35441$ & $\pm 0.37$   & ---   & --- \\
$Y_{min,2}$ & $0.09644$ & $\bullet$   & ---   & $\bullet$ \\
$Y_{max,2}$ & $0.34904$ & $\bullet$   & ---   & $\bullet$ \\
$R_{in,2}$ & $\pm 0.19640$ & $\pm 0.19$   & ---   & --- \\ 
$R_{out,2}$ & $\pm 0.37366$ & $\pm 0.37$   & ---   & --- \\ \hline
\end{tabular}}
\caption{Comparison of perturbation theory and numerics for $2\!:\!1$ 
resonances ($\kappa = 0.75$, $\delta = 0.2$).  The quantities computed are 
defined in the caption of Table \ref{primary1}.} \label{primary2}
\end{table}

\subsection{Secondary Resonances}

Our comparison between theory and numerics for secondary resonances is 
summarized in Table \ref{secondary}.

\begin{table}[t] 
\begin{tabular}{l r}
$\hspace{4.3 cm} \epsilon = 0.01$ \hspace{2.9 cm} & \hspace{0.6 cm} 
$\epsilon = 0.05$ \\ 
\end{tabular}
\\
\centerline{
\begin{tabular}{|c||c|c||c|r|} \hline
Quantity & Perturbative & Numerical & Perturbative & Numerical \\ \hline
$Y_2$ & $0.37358$ & $\bullet$   & $0.37358$   & $\bullet$ \\
$Y_s$ & $0.37294$ & $\bullet$   & $0.37036$   & $\bullet$ \\
$R_s$ & $\pm 0.86364$ & $\pm 0.88$ & $\pm 0.86065$  & $\pm 0.88$ \\
$S_s$ & $\pm 0.68389$ & $\pm 0.687$ & $\pm 0.68293$  & $\pm 0.68$ \\
$Y_c$ & $0.37422$   & $\bullet$   & $0.37680$ & $\bullet$ \\
$(R_c,S_c)$ & See text & $(\pm 0.691, \pm 0.332)$   & See text  & $(\pm 0.697, \pm 0.330)$ \\
$Y_{min}$ & $0.34814$ & $\bullet$   & $0.31670$   & $\bullet$ \\
$Y_{max}$ & $0.39902$   & $\bullet$   & $0.43046$ & $\bullet$ \\
$Y_{min,2}$ & $0.35571$ & $\bullet$   & $0.32989$   & $\bullet$ \\
$Y_{max,2}$ & $0.41237$ & $\bullet$   & $0.46240$   & $\bullet$ \\ \hline
\end{tabular}}
\caption{Comparison of perturbation theory and numerics for $4\!:\!1$ 
resonances ($\kappa = 2.5$, $\delta = 1$).  Tha action value of the secondary
 resonance is denoted $Y_2$.  Saddles that intersect the $R$-axis are denoted
 $(R_s,0)$, and those that intersect the $S$-axis are denoted $(0,S_s)$.  
Centers are denoted $(R_c,S_c)$.  The other quantities computed are defined 
in the caption of Table \ref{primary1}.} \label{secondary}
\end{table}

\begin{figure}
        \begin{centering}
                \includegraphics[width = 2.5 in, height = 3 in]
{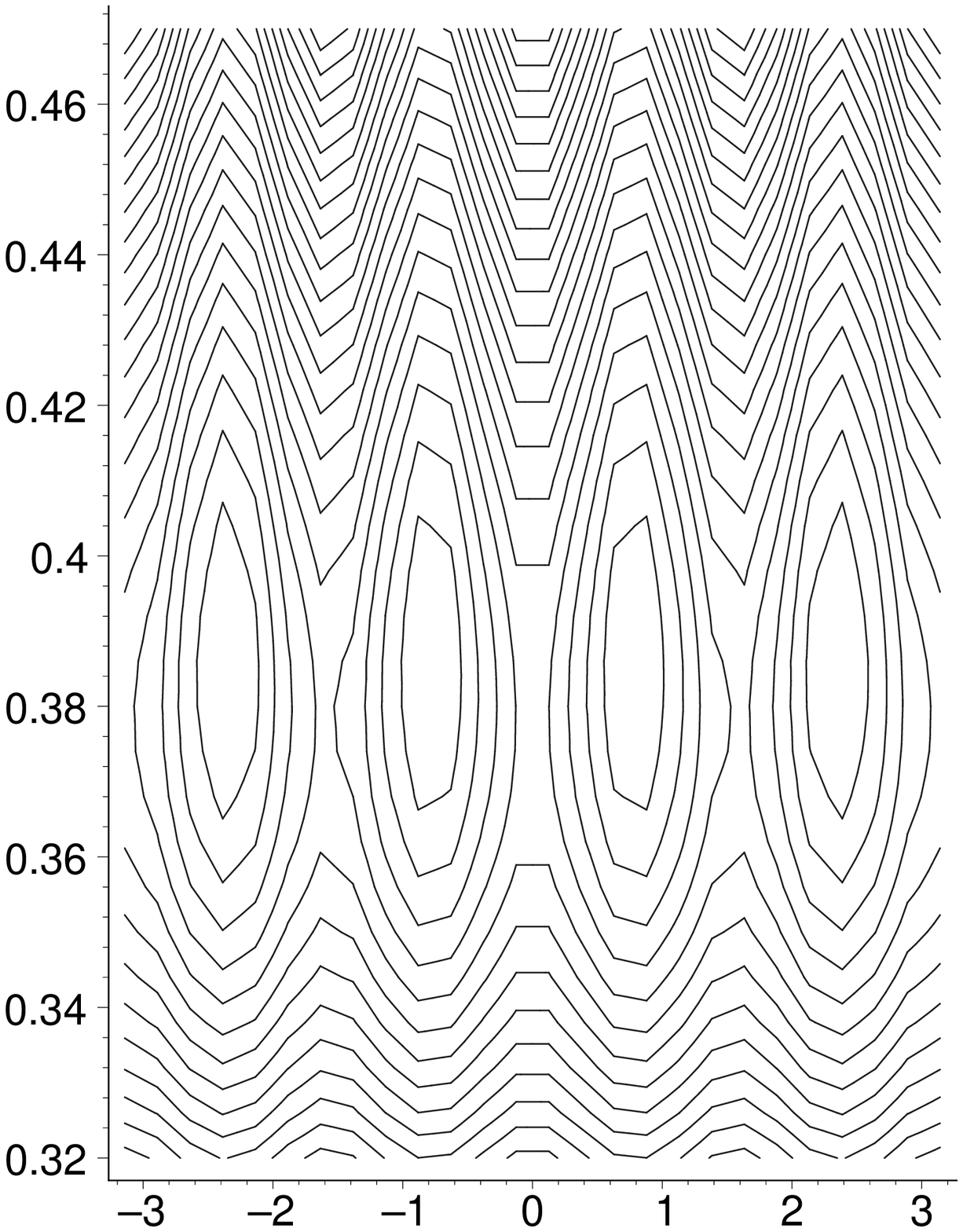}

                \caption{Resonance Hamiltonian $K_2$ for $\kappa = 2.5$, 
$\delta = 1$, and $\epsilon = 0.05$.} \label{k2}
        \end{centering}
\end{figure}

\begin{figure}
                \centerline{
                (a)
                \includegraphics[width=0.40\textwidth, height=0.50\textwidth]
{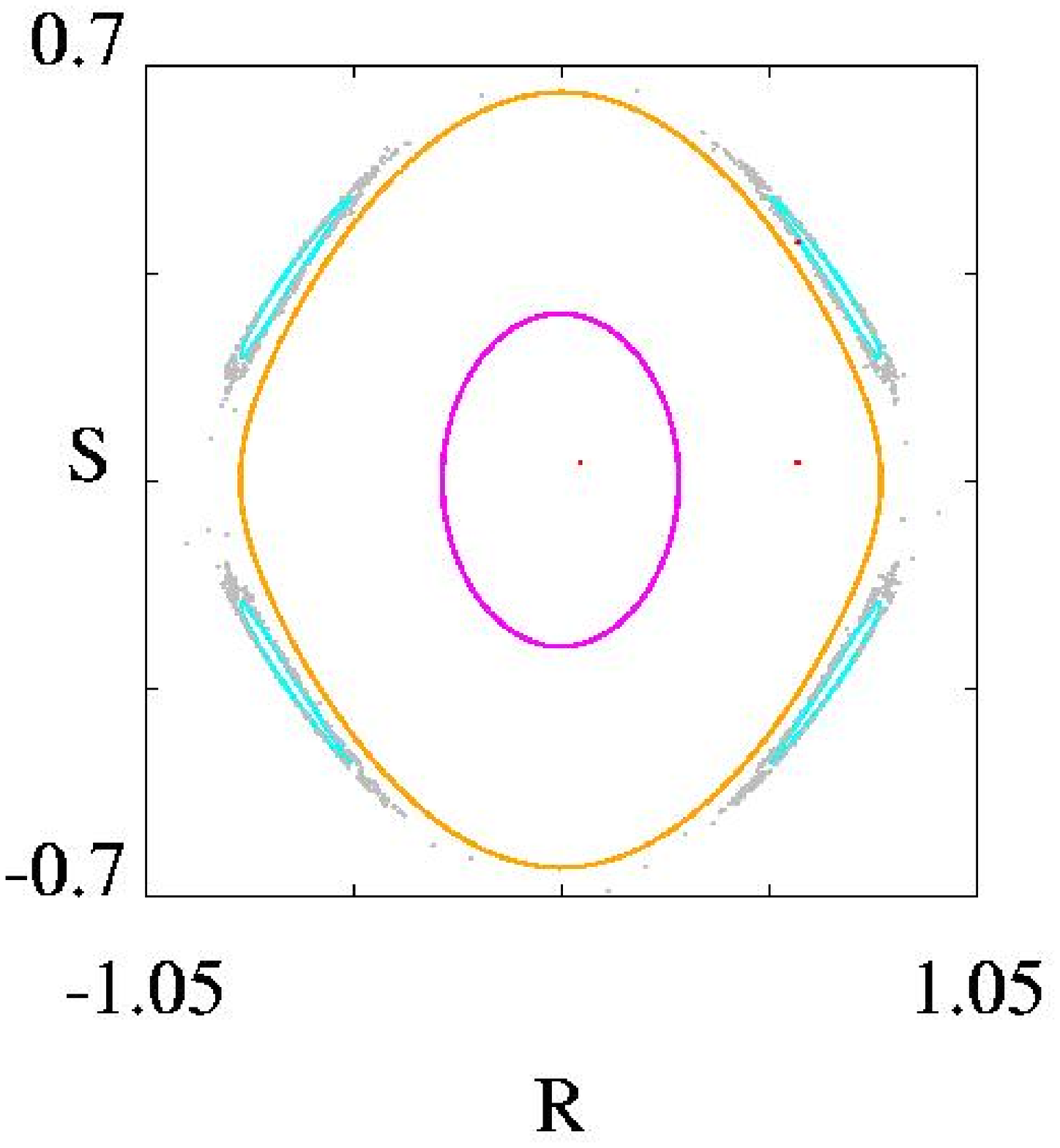}
                \hspace{1 cm}
                (b)
                \includegraphics[width=0.40\textwidth, height=0.50\textwidth]
{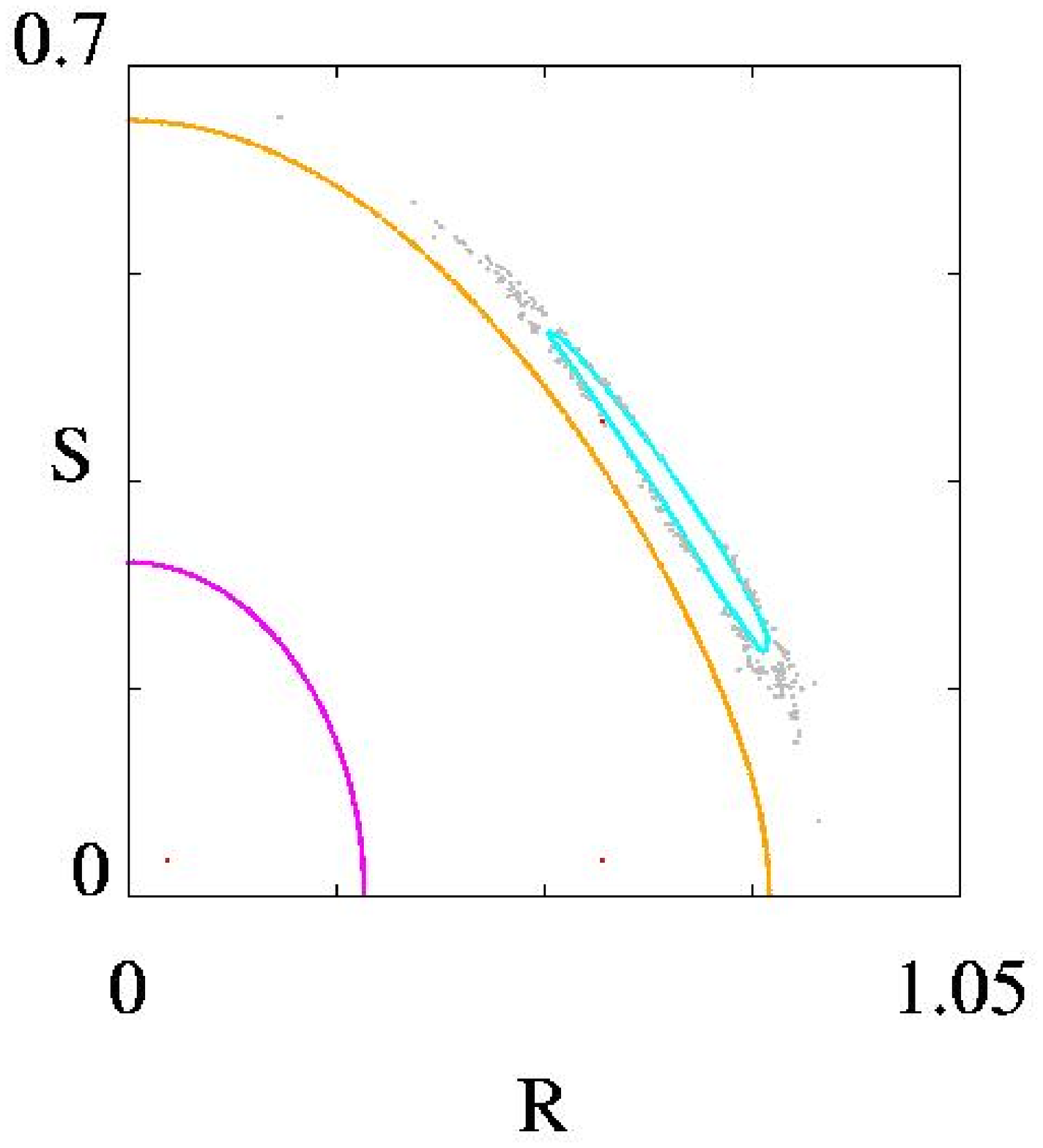}}


                \caption{(a) Poincar\'e section for $\kappa = 2.5$, 
$\delta = 1$, and $\epsilon = 0.05$.  Note that there is no $2\!:\!1$ 
resonance band for this choice of $(\kappa,\delta)$.  The $4\!:\!1$ resonance
 is depicted. (b) Upper right corner of (a).} \label{res22}

\end{figure}

We study $4\!:\!1$ resonances for $\kappa = 2.5$ and 
$\delta = 1$.  No $2\!:\!1$ resonances exist for this choice of parameters.
  The resonance Hamiltonian is depicted for $\epsilon = 0.05$ in Figure 
\ref{k2}.  The corresponding Poincar\'e section is shown in Figure 
\ref{res22}.  

When $\epsilon = 0.01$, we observe numerically that centers are located at 
approximately $(R,S) = (\pm 0.691, \pm 0.332)$.  With $R = \pm 0.691$, we 
predict a value of $S = \pm 0.32384$.  With $S = \pm 0.332$, we predict a 
value of $R = \pm 0.68362$.  These predictions are remarkably good, as we 
have used leading-order perturbation theory to derive analytical predictions 
for $4\!:\!1$ (secondary) resonances.  However, they are not as good as those
 obtained for the location of saddles in this case or the location of centers
 for $2\!:\!1$ (primary) resonances.

When $\epsilon = 0.05$, numerical simulations suggest that centers are 
located at about $(R,S) = (\pm 0.697, \pm 0.330)$.  Using $R = \pm 0.697$ 
leads to a prediction of $S = \pm 0.31912$.  Using $S = \pm 0.330$ leads to a
 prediction of $R = \pm 0.68721$.

\subsection{Tertiary Resonances}

Our comparison between theory and numerics for tertiary resonances is 
summarized in Table \ref{tertiary}.

\begin{table}[t] 
\centerline{
\begin{tabular}{|c||c|r|} \hline
Quantity & Perturbative & Numerical \\ \hline
$Y_3$ & $0.36857$ & $\bullet$   \\
$Y_s$ & $0.36851$ & $\bullet$   \\
$R_s$ & $\pm 0.85850$ & $\pm (0.859-0.860)$ \\
$Y_c$ & $0.36863$   & $\bullet$  \\
$R_c$ & $\pm 0.85864$ & $\pm 0.870$   \\
$Y_{min}$ & $0.36214$ &  $\bullet$  \\
$Y_{max}$ & $0.37500$   & $\bullet$   \\
$Y_{min,2}$ & $0.36614$ & $\bullet$  \\
$Y_{max,2}$ & $0.38653$ & $\bullet$   \\ \hline
\end{tabular}}
\caption{Comparison of perturbation theory and numerics for $6\!:\!1$ 
resonances ($\kappa = 3.8$, $\delta = 1$, $\epsilon = 0.01$).  Tha action 
value of the tertiary resonance is denoted $Y_3$.  The other quantities 
computed are defined in the caption of Table \ref{primary1}.} 
\label{tertiary}
\end{table}

\begin{figure}
        \begin{centering}
                \includegraphics[width = 2.5 in, height = 3 in]
{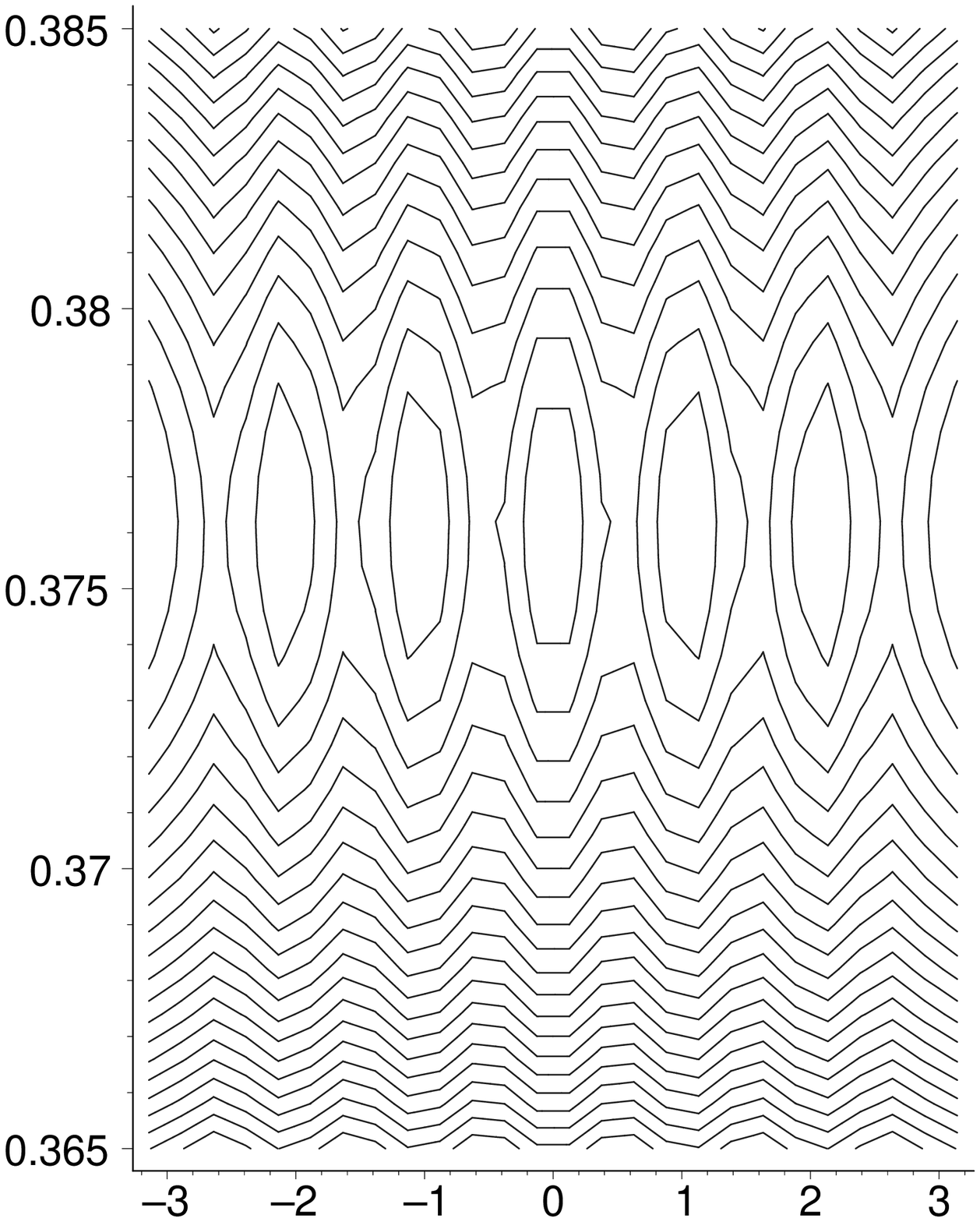}

                \caption{Resonance Hamiltonian $K_3$ for $\kappa = 3.8$, 
$\delta = 1$, and $\epsilon = 0.01$.} \label{k301}
        \end{centering}
\end{figure}

\begin{figure}
                \centerline{
                (a)
                \includegraphics[width=0.40\textwidth, height=0.50\textwidth]
{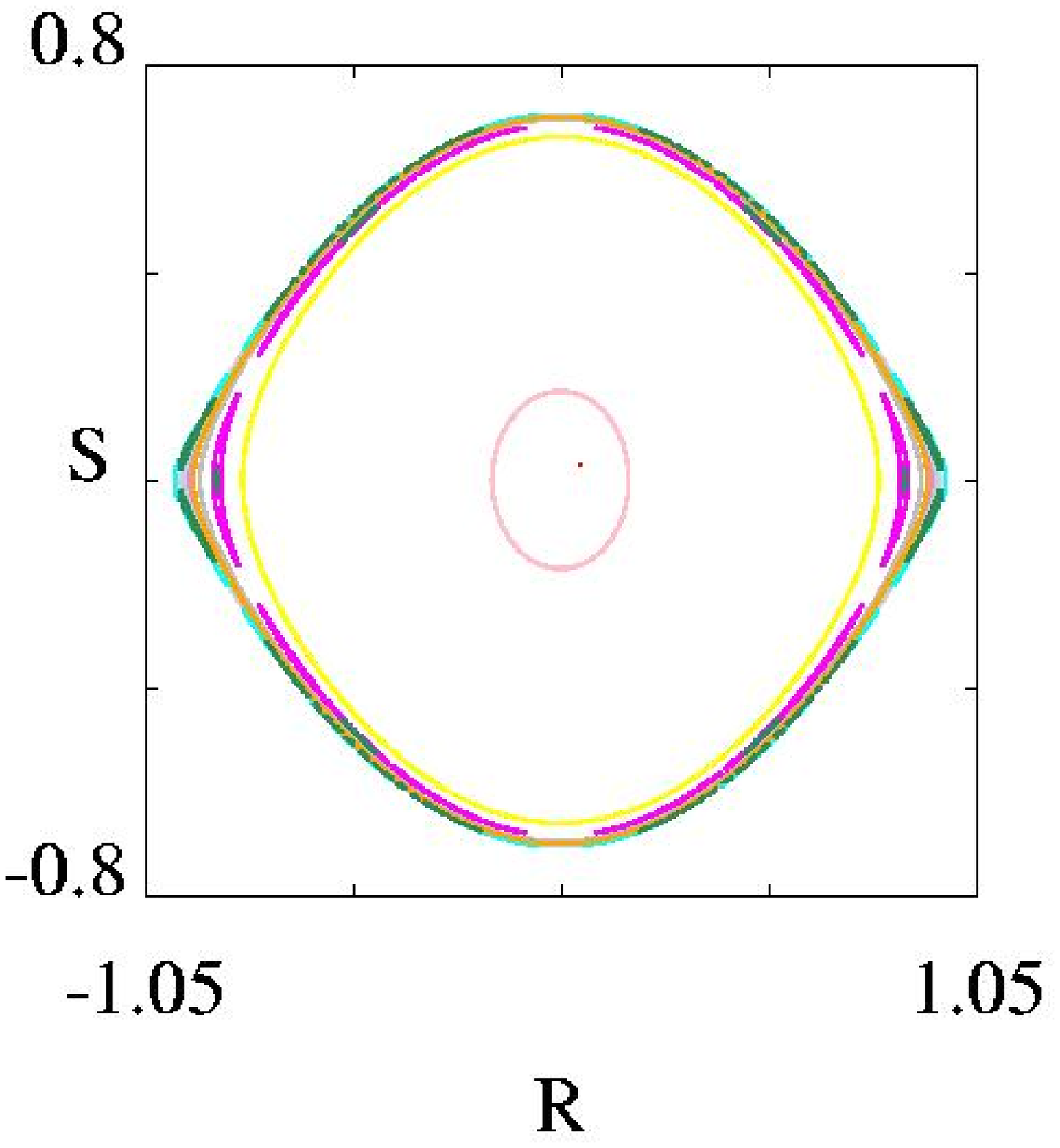}
                \hspace{1 cm}
                (b)
                \includegraphics[width=0.40\textwidth, height=0.50\textwidth]
{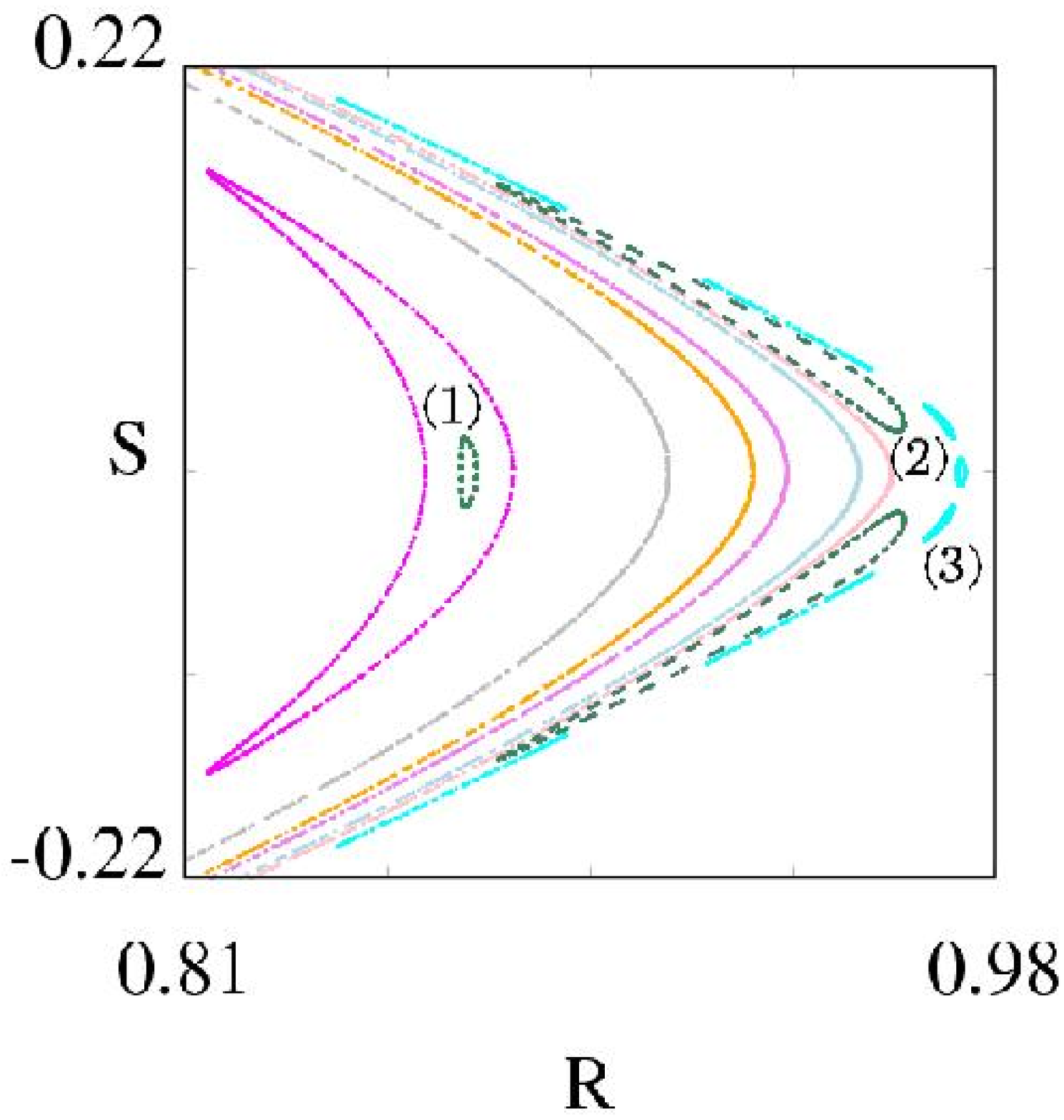}}


                \caption{(a) Poincar\'e section for $\kappa = 3.8$, 
$\delta = 1$, and $\epsilon = 0.01$.  (b) Close-up of the resonances in (a). 
Both $6\!:\!1$ (1) and $8\!:\!1$ (2) resonances are displayed.  A 
higher-order resonance (3) is also depicted.  Although not shown, $4\!:\!1$ 
resonances are also present for this choice of parameter values.} \label{res3}

\end{figure}

We consider $6\!:\!1$ resonances for $\kappa = 3.8$, $\delta = 1$, and 
$\epsilon = 0.01$.  No $2\!:\!1$ resonances exist for this choice of 
parameters, but $4\!:\!1$ resonances do exist.  The resonance Hamiltonian is 
depicted for $\epsilon = 0.01$ in Figure \ref{k301}.  The corresponding 
Poincar\'e section is shown in Figure \ref{res3}.  (The $4\!:\!1$ resonance 
bands are not shown in this plot.)  

The predictions for centers are not as good as those for saddles, but there 
is nevertheless good quantitative agreement between observation and 
prediction, especially considering that a leading-order perturbation method 
has been employed.  Of course, given that higher-order resonances occupy 
smaller regions of phase space, the absolute errors indicate that these 
predictions are not as good as the same absolute errors would be when 
studying lower-order resonances.  This caveat notwithstanding, our 
theoretical analysis does an excellent job of determining the location of 
resonances and offers a useful tool for locating high-order resonances (and 
thus studying band structure in great detail) in numerical simulations.

\section{Conclusions}

In this paper, we studied in depth the band structure of BECs in 
periodic lattices.  We approached this problem using a coherent structure 
ansatz, in contrast to the Bloch wave ansatz of earlier 
studies.\cite{wu4,machholm,diak}

Using a technically delicate perturbative approach relying on elliptic 
function solutions of the integrable NLS, we examined the spatial resonance 
structure (band structure) of coherent structure solutions of the NLS in 
considerable detail, providing both an analytical description and numerical 
verifications of this theory.  We derived conditions for the onset of 
$2m'\!:\!1$ spatial resonances for all integer $m'$ and developed analytical 
expressions for the width of these resonance bands and the locations of 
saddles and centers therein.  Comparison with numerical simulations of 
primary, secondary, and tertiary resonances illustrate the applicability of 
our analytical theory.  

Utilizing a simpler perturbative approach that employs Lindstedt's method and 
multiple scale analysis, we also established wave number-amplitude relations 
for coherent structure solutions of the NLS with a periodic potential.  In so
 doing, we explored $2\!:\!1$ spatial resonances and illustrated the utility 
of phase space analysis for the study of band structure as well as the 
structure of modulated amplitude waves in BECs.  

In sum, our perturbative approach does an excellent job of determining the 
location of resonances and analyzing their structure and offers a useful tool
 for locating high-order resonances (and thus studying BEC band structure in 
great detail) in numerical simulations.  An important open direction, to be 
addressed in a future publication, is the extent to which the theory 
developed here is an effective starting point for studies of the chaotic 
dynamics of BECs.

\section*{Acknowledgements}

Valuable conversations with Eric Braaten, Michael Chapman, Mark Edwards, 
Nicolas Garnier, Brian Kennedy, Yueheng Lan, Igor Mezi\'c, Peter Mucha, and 
Dan Stamper-Kurn are gratefully acknowledged.  We are especially grateful to 
Jared Bronski, Richard Rand, and Li You for several extensive discussions 
concerning this project and to Panos Kevrekidis, Boris Malomed, Alexandru 
Nicolin, and an anonymous referee for critically reading and offering useful 
suggestions that greatly improved this manuscript.


\end{document}